\documentclass[fleqn,usenatbib]{mnras}
\usepackage{newtxtext,newtxmath}
\usepackage[T1]{fontenc}

\DeclareRobustCommand{\VAN}[3]{#2}
\let\VANthebibliography\thebibliography
\def\thebibliography{\DeclareRobustCommand{\VAN}[3]{##3}\VANthebibliography}

\usepackage{graphicx}	% Including figure files
\usepackage{amsmath}	% Advanced maths commands

\newcommand*  {\diff}       {\mathop{}\!\mathrm{d}}

\newcommand*  {\Sect}[1]    {Section~\ref{#1}}
\newcommand*  {\App}[1]     {Appendix~\ref{#1}}
\renewcommand*{\vec}[1]     {\boldsymbol{#1}}
\newcommand*  {\uvec}[1]	{\hat{\vec{#1}}}

\newcommand*  {\p}          {\partial}

\newcommand*  {\pdiff}[2]   {\frac{\p{#1}}{\p{#2}}}
\newcommand*  {\tdiff}[2]   {\frac{\diff{#1}}{\diff{#2}}}

\newcommand*  {\cross}	    {\times}

\newcommand*  {\sub}[2]     {{#1}_{\mathrm{#2}}}

\newcommand*  {\kms}        {{\mathrm{km\,s^{-1}}}}

\title[The cross-section for ISO capture]
{Capture of interstellar objects I: the capture cross-section}

\author[Dehnen, Hands]{%
Walter~Dehnen$^{\!\!1,2,3}$ and
Thomas~O.~Hands$^{4}$
\smallskip
\\
$^1$ Astronomisches Recheninstitut, Zentrum f{\"u}r Astronomie der Universit{\"a}t Heidelberg, M{\"o}nchhofstra\ss{}e. 12-14, 69120, Heidelberg, Germany\\
$^2$ Universit{\"a}ts-Sternwarte M{\"u}nchen, Scheinerstra\ss{}e 1, 81679, M{\"u}nchen, Germany\\
$^3$ School for Physics and Astronomy, University of Leicester, University Road, LE1 7RH, UK\\
$^4$ Institut f{\"u}r Computergest{\"u}tzte Wissenschaften, Universit{\"a}t Z{\"u}rich, Winterthurerstrasse 190, CH-8057 Z{\"u}rich, Switzerland
}

\date{Accepted XXX. Received YYY; in original form ZZZ}

\pubyear{2021}

\begin{document}

\defcitealias{paper2}{paper~2}
%%%%%%%%%%%%%%%%%%%%%%%%%%%%%%%%%%%%%%%%%%%%%%%%%%%%%%%
\label{firstpage}
\pagerange{\pageref{firstpage}--\pageref{lastpage}}
\maketitle

\begin{abstract}
We study the capture of interstellar objects (ISOs) by a planet-star binary with mass ratio $q\ll1$, semi-major axis $\sub{a}p$, orbital speed $\sub{v}c$, and eccentricity $\sub{e}p$. Very close (slingshot) and wide encounters with the planet are amenable to analytic treatment, while numerically obtained capture cross-sections $\sigma$ closely follow the analytical results even in the intermediate regime. Wide interactions can only generate energy changes $\Delta E\lesssim q\sub{v}c^2$, when $\sigma\propto v_\infty^{-2} |\ln\Delta E|^{2/3}$ (with $v_\infty$ the ISO's incoming speed far away from the binary), which is slightly enhanced for $\sub{e}p>0$. Energy changes $\Delta E\gtrsim q\sub{v}c^2$, on the other hand, require close interactions when $\sigma\propto (v_\infty \Delta E)^{-2}$ hardly depending on $\sub{e}p$. Finally, at $\Delta E\gtrsim \sub{v}c^2$, the cross-section drops to zero, depending on the planet's radius $\sub{R}p$ through the Safronov number $\Theta=q\sub{a}p/\!\sub{R}p$. We also derive the cross-sections for collisions of ISOs with planets or moons.
\end{abstract}


%%%%%%%%%%%%%%%%%%%%%%%%%%%%%%%%%%%%%%%%%%%%%%%%%%%%%%%
\section{Introduction}
%%%%%%%%%%%%%%%%%%%%%%%%%%%%%%%%%%%%%%%%%%%%%%%%%%%%%%%
The recent discovery of the first two known interstellar objects (ISOs) to visit the Solar system -- 1I/`Oumuamua \citep[][]{Meech2017,ISSI2019}  and 2I/Borisov \citep{JewittLuu2019} -- has raised many more questions than it has answered. The likely origin for such objects is that they form around alien stars just as asteroids and comets in the Solar system -- as a part of the planet formation process -- and are left as debris when this process is completed. They are then ejected from their home system by some dynamical event -- be it a stellar fly-by/intra-cluster interaction \citep[see e.g.,][]{HandsEtAl2019} or interaction with a giant planet \citep[see e.g.,][]{Raymond2018b} -- and travel through interstellar space until they have a chance encounter with another star. This unique origin has led to a great deal of interest from the planet formation community. Studying ISOs that transit through the Solar system affords us for the first time the opportunity to glimpse material and therefore chemistry directly from the planet formation environment around distant stars. There have been multiple fruitful Earth-based observations of transient ISOs \citep[see e.g.,][for a selection of `Oumuamua observations]{Jewitt2017,Bannister2017,TrillingEtAl2018}, as well as suggestions of future sample return missions \citep{Hein2019}. However, with the only two known ISOs currently moving away from the Solar system, chances to study them remain few and far between.

The identification of a population of unequivocally interstellar objects residing within the Solar system would enable study on a longer time-scale than waiting for a chance encounter. Due to the time-reversible nature of Hamiltonian dynamics, it is clear that if ISOs can be ejected by giant planets in their host systems, they can also be captured by the giant planets in the Solar system. The purpose of this study is to systematically investigate, both analytically and numerically, the cross-section for capturing ISOs by a planet-star binary in general and the Solar system in particular. Previous analytical studies \citep{Radzievskii1967, BandermannWolstencroft1970, Heggie1975, RadzievskiiTomanov1977, PineaultDuquet1993} were limited to to close interactions with a planet on a circular orbit. Here, we extend this previous work to non-circular orbits and also consider wider interactions, which are relevant for capturing ISOs at very small incoming asymptotic speeds $v_\infty$ and/or onto orbits with large semi-major axis. Previous numerical studies \citep{ValtonenInnanen1982, HandsDehnen2020, NapierAdamsBatygin2021} generally focused on the Solar system and used a rather limited number of simulated trajectories. Here, we extend these to numerically derive the capture cross-section $\sigma$ as function of the incoming asymptotic speed $v_\infty$ of the ISO, the semi-major axis $a$ of the orbit onto which the ISO is captured, the planet-to-star mass ratio $q$, and the eccentricity $\sub{e}p$ of the planet's orbit. 

This paper is organised as follows. In \Sect{sec:analytic} we use analytic approximations to obtain the capture cross-section in the limits of close (slingshot) and very wide encounters of the ISO with the planet. Then, in \Sect{sec:sigma:sim} we perform numerical simulations to obtain the capture cross-section  for planets of various masses and orbital eccentricities. \Sect{sec:coll+eject} briefly discusses two related processes: the collision of ISOs with Solar system planets and the ejection of bound objects by exoplanets. Finally, \Sect{sec:conclude} summarises and concludes this paper, while in an accompanying paper \citepalias[\citealt{paper2}, hereafter][]{paper2}, we use the cross-section calculated here to evaluate the capture rate and the population of captured ISO with particular emphasis on the Solar system.
 
%%%%%%%%%%%%%%%%%%%%%%%%%%%%%%%%%%%%%%%%%%%%%%%%%%%%%%%
\section{Analytic treatment of ISO capture}
\label{sec:analytic}
We consider a planet-star binary with masses $\sub{m}p\ll\sub{m}s$ encountering a test particle, hereafter \emph{interstellar object} (ISO), with incoming asymptotic speed $v_\infty$. In this section, we summarise and derive analytic results regarding collision with the planet (\S\ref{sec:analytic:collide}) as well as capture by a close (\S\ref{sec:analytic:capture:strong}) or by wide (\S\ref{sec:analytic:capture:weak}) encounter with it. Table~\ref{tab:symbols} summarises most symbols used.

For collisions and close encounters with the planet, we use the \emph{impulse approximation}, which models the ISO orbit as a barycentric hyperbola instantaneously deflected by the planet. In \S\ref{sec:analytic:capture:weak} we use perturbation theory to shed light on the behaviour at very wide encounters, when the ISO passes the planet at distances comparable or even larger than its semi-major axis. In the intermediate regime analytic insight is limited, but our numerical results in the next section indicate that the actual capture cross-section is well described by either of these limiting cases.

%%%%%%%%%%%%%%%%%%%%%%%%%
\begin{table}
	\caption{Summary of most symbols used. \label{tab:symbols}}
	\begin{tabular}{@{}l@{\hspace{2ex}}p{7.05cm}@{}}
		symbol & meaning \\
		\hline
		$\sub{m}s,\,\sub{m}p$
			& masses of star and planet
		\\
		$M,\,q$
			& 	$\sub{m}s+\sub{m}p$ (total binary mass),
				$\sub{m}p/\sub{m}s$ (binary mass ratio)
		\\
		$\sub{R}p,\,\sub{v}{esc,p}$
			& planet's radius, $\sqrt{2G\sub{m}p/\sub{R}p}$ (escape speed from its surface)
		\\
		$\sub{R}s,\,\sub{v}{esc,s}$
			& stellar radius, $\sqrt{2G\sub{m}s/\sub{R}s}$ (escape speed from its surface)
		\\
		$\sub{a}p,\,\sub{e}p$
			&  planet's semi-major axis and eccentricity
		\\
		$\sub{\eta}p,\,\sub{\ell}p$
			&  planet's eccentric and mean anomaly
		\\
		$\sub{\Omega}p$
			&  planet's mean motion $\sqrt{GM/\smash{\sub{a}{p}^3}}$
		\\
		$\sub{r}p,\,\sub{\vec{v}}p$
			&  planet's barycentric orbital radius and velocity (equation~\ref{eq:r,v})
		\\
		$\sub{v}c,\,\sub{v}r$
			& $\sqrt{GM/\sub{a}p},\,\sqrt{GM/\sub{r}p}$ (equation~\ref{eq:v:r})
		\\
		$\sub{n}{iso},\,n_1$
			& number density of ISOs in interstellar space and near planet
		\\
		$a,\,\sub{v}a$
			& semi-major axis of captured ISO, $\sqrt{GM/a}$
		\\
		$v_\infty$ %, $V$
			& asymptotic barycentric speed of incoming ISO %, $[v_\infty^2+\sub{v}a^2]^{1/2}$
		\\
		$X$
		    & $(v_\infty^2+\sub{v}a^2)/q\sub{v}c^2$: dimensionless measure of energy change
		\\
		$\vec{v}_{1,2}$
			& ISO's barycentric velocity just before and just after deflection
		\\
		$\vec{w}_{1,2}$
			& $\vec{v}_{1,2}-\sub{\vec{v}}p$: planetocentric ISO velocity
		\\
		$\alpha,\,\theta$
			&	angle between $\sub{\vec{v}}p$ and $\vec{v}_1$,
				angle between $\sub{\vec{v}}p$ and $\vec{w}_1$
		\\
		$C,\,D$
			&	$2\sub{v}r^2 - \sub{v}p^2 - w^2 + v_\infty^2$, $-2\sub{v}r^2 + \sub{v}p^2 + w^2 + \sub{v}a^2$
			(equations~\ref{eq:C} and~\ref{eq:D})
		\\
		$b,\,\psi$
			& ISO's barycentric impact parameter and impact angle
		\\
		$\sub{b}p,\,\sub{\psi}p$
			& ISO's planetocentric impact parameter and impact angle
		\\
		$b_0,\,\rho$
			& offset and radius of planetocentric impact disc
				(equations~\ref{eqs:b0,rho})
		\\
		$\sub{d}s,\,\sub{d}p$
			& ISO's closest-approach distances to star and planet
		\\
		$s,\,\sub{s}p$
		    & barycentric and planetocentric
		    semi-latus rectum of ISO
		\\ 
		$\Theta$
			& $G\sub{m}p/\sub{R}p\sub{v}p^2$ (Safronov number)
		\\
		\hline		
	\end{tabular}
\end{table}
%%%%%%%%%%%%%%%%%%%%%%%%%
In the impulse approximation, the deflection by the planet is itself modelled as that between incoming and outgoing asymptote of an hyperbolic planetocentric orbit. If the peri-centre of this orbit is within the planet, the ISO is not deflected but collides with the planet. This limits the ability of large planets to inflict strong deflections and hence to capture ISOs with large $v_\infty$ or onto strongly bound orbits.

The impulse approximation accounts for the gravity of the planet twice: once as part of the barycentric total mass $M\equiv\sub{m}s+\sub{m}p$ and once for calculating the deflection. The error made by this is $O(q)$ with the mass ratio $q\equiv \sub{m}p/\sub{m}s$ and cannot be neglected in applications to binary stars (even though this has been done). However, for planet-star systems considered here $q\ll1$ and we can safely neglect this error. In the following we shall replace $1+q\to1$ and use `$\doteq$' instead of `$=$' to denote relations that are exact up to this approximation.

\begin{figure}
    \begin{center}
	    \includegraphics[width=\columnwidth]{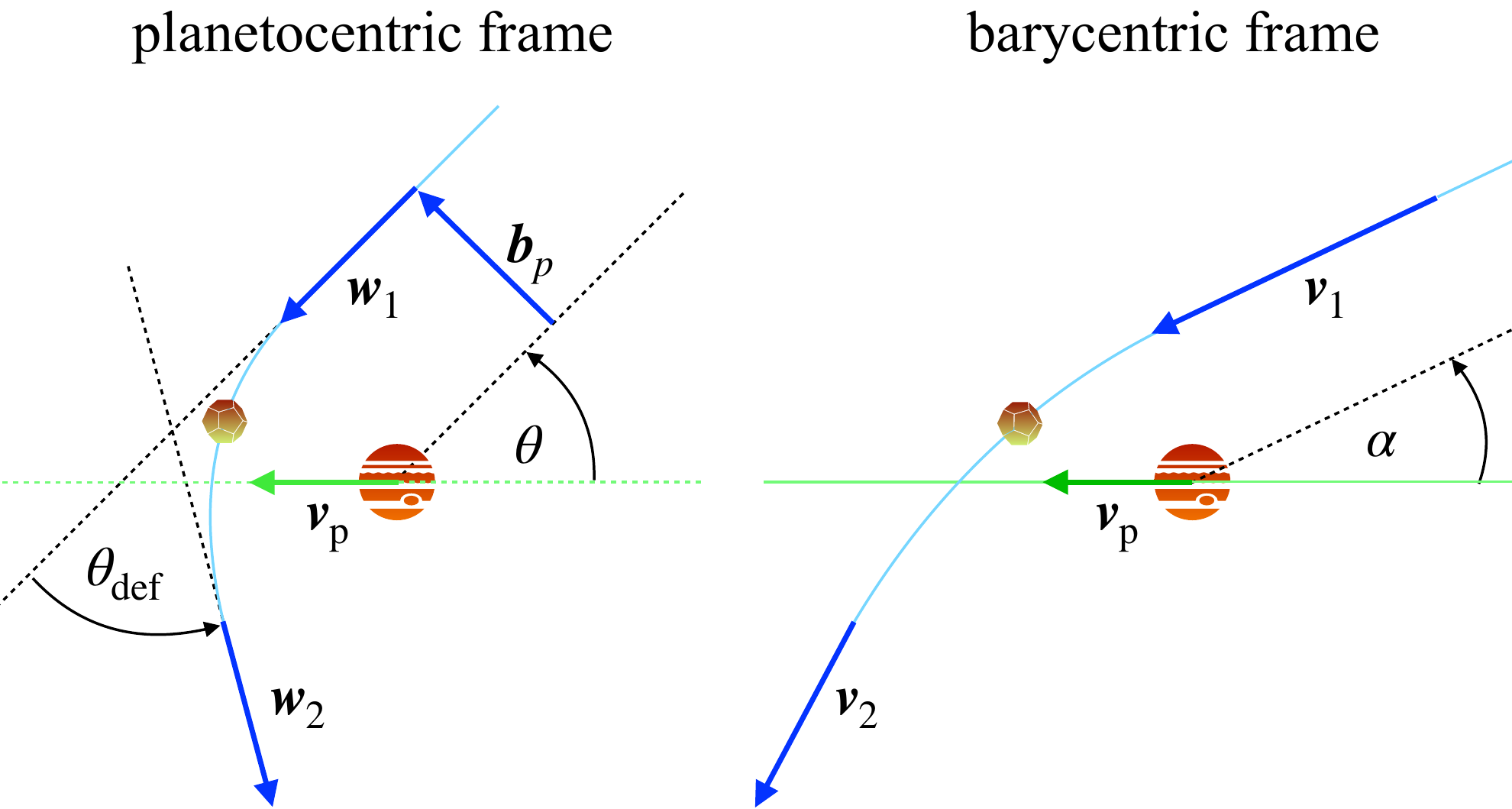}
    \end{center}
	\caption{\label{fig:Orbit}
	Swing-by of the ISO by the planet. In the frame moving with the planet (left) the ISO is deflected by angle $\sub{\theta}{def}$ but retains its speed ($|\vec{w}_1|=|\vec{w}_2|$). If the ISO passes in front of the planet, the outgoing barycentric speed $|\vec{v}_2|$ (and hence energy) is smaller than the incoming speed $|\vec{v}_1|$, enabling capture. The depicted example trajectory has $\sub{\psi}p=0$ in equation~\eqref{eq:vp:impact} when the vectors $\sub{\vec{v}}p$, $\vec{w}_1$, and $\sub{\vec{b}}p$ are co-planar.
	}
\end{figure}

Let $\sub{a}p$, $\sub{e}p$ and $\sub{\eta}p$ denote, respectively, the planet's semi-major axis, eccentricity, and eccentric anomaly at the moment of the impulsive encounter with the planet. Then the planet's barycentric radius and speed at that moment are, respectively,
\begin{align}
	\label{eq:r,v}
	\sub{r}p \doteq (1-\sub{e}p\cos\sub{\eta}p)\,\sub{a}p
	\quad\text{and}\quad
	\sub{v}p \doteq \sqrt{\frac	{1+\sub{e}p\cos\sub{\eta}p}
	{1-\sub{e}p\cos\sub{\eta}p}}\,
	\sub{v}c
\end{align}
with $\sub{v}c\equiv\sqrt{GM/\sub{a}p}\doteq\sqrt{G\sub{m}s/\sub{a}p}$. The speed of the ISO when it enters the planet's sphere of influence (which is approximated to be of negligible size) follows from energy conservation as
\begin{align}
	\label{eq:v:1}
	v_1^2 = 2\sub{v}r^2 + v_\infty^2,
\end{align}
where
\begin{align}
	\label{eq:v:r}
	\sub{v}r^2	\equiv \frac{GM}{\sub{r}p}
	\doteq \frac{\sub{v}c^2}{1-\sub{e}p\cos\sub{\eta}p} .
\end{align}
In the frame of the planet, the ISO has incoming and outgoing velocities \begin{align}
    \label{eq:w:12}
    \vec{w}_{1,2}\equiv\vec{v}_{1,2}-\sub{\vec{v}}p
\end{align}
and speed $w\equiv|\vec{w}_2| = |\vec{w}_1|$ satisfying
\begin{align}
	\label{eq:w:alpha}
	w^2 = \sub{v}p^2 + v_1^2 - 2\sub{v}p v_1^{} \cos\alpha,
\end{align}
where $\alpha$ is the angle between the ISO's and planet's barycentric velocities just before the encounter, see also \autoref{fig:Orbit}.

%%%%%%%%%%%%%%%%%%%%%%%%%%%%%%%%%%%%%%%%%%%%%%%%%%%%%%%
\subsection{Cross-section for collision with the planet or a moon}
\label{sec:analytic:collide}
Collisions are important in the context of captures, as they compete with those captures that require large energy changes, see also \Sect{eq:analytic:strong:collide}. Therefore, first we derive the cross-section for collisions of ISOs with the planet.

If $\sub{b}p$ is the impact parameter of the ISO's hyperbolic planetocentric orbit, then its distance of closest approach to the planet is
\begin{align}
	\label{eq:dp}
	\sub{d}p = \frac{1}{w^2}
	\left(\sqrt{\sub{b}p^2w^4 + G^2\sub{m}p^2} - G\sub{m}p\right)
\end{align}
(see also equation~\ref{eq:closest:r}). The ISO will collide if $\sub{d}p\le\sub{R}p$, the radius of the planet. Solving~\eqref{eq:dp} for $\sub{b}p$ at $\sub{d}p=\sub{R}p$ gives the local cross-section
\begin{align}
	\label{eq:sigma:collision:p}
	\sub{\sigma}p = \pi \sub{b}p^2
			= \pi \sub{R}p^2 \left[1+\frac{\sub{v}{esc,p}^2}{w^2}\right]
\end{align}
for collisions of ISOs with the planet. Here, $\sub{v}{esc,p}\equiv\sqrt{2G\sub{m}p/\sub{R}p}$ is the escape speed from the surface of the planet. If $n_1$ is the number density of ISOs at the planet, the flux of colliding ISOs is $F=n_1w\sub{\sigma}p$ and depends on the relative orientation $\alpha$. From equation~\eqref{eq:w:alpha},
\begin{align}
	\label{eq:dcosalpha}
	\diff\cos\alpha=-\frac{w}{v_1\sub{v}p}\diff w
\end{align} and averaging over all directions (assuming the distribution of $\vec{v}_1$ is isotropic, which holds if the distribution of $\vec{v}_\infty$ is isotropic) obtains
\begin{align}
	\langle F \rangle_\alpha = \frac{n_1\pi\sub{R}p^2}{v_1}
	\left(v_1^2+\tfrac13\sub{v}p^2+\sub{v}{esc,p}^2\right).
\end{align}
Next, we must average the flux over the eccentric planet orbit:
\begin{align}
	\label{eq:F:ave:orbit}
	\langle F \rangle &\equiv \frac{1}{2\pi} \int_0^{2\pi} \langle F \rangle_\alpha
	(1-\sub{e}p\cos\sub{\eta}p) \diff\sub{\eta}p
	\\
	&\doteq \frac{\sub{n}{iso} \pi \sub{R}p^2}{v_\infty} \left(v_\infty^2 +\sub{v}{esc,p}^2 +\tfrac73\sub{v}c^2\right),
\end{align}
where we have used equations~(\ref{eq:r,v}-\ref{eq:v:r}) and exploited that for a mono-energetic population of ISOs $n\propto\int\delta(E-\tfrac12v_\infty^2)\diff^3\vec{v}\propto v$, such that $n_1/v_1=\sub{n}{iso}/v_\infty$\footnote{$\sub{n}{iso}$ is the number density of ISOs in the Solar vicinity, but far enough to not be enhanced by gravitational focussing. Current estimates for $\sub{n}{iso}$ range between 0.01 and 0.2 per au$^3$, depending on object type and size, see also the introduction of \citetalias{paper2}.}. Finally, the cross-section for collisions with the planet follows as
\begin{align}
	\label{eq:sigma:collision}
	\sub{\sigma}{coll,p} = \frac{\langle F\rangle}{\sub{n}{iso} v_\infty}
		\doteq \pi\sub{R}p^2
		\left[1 + \frac{\sub{v}{esc,p}^2}{v_\infty^2} + \frac73\frac{\sub{v}c^2}{v_\infty^2} \right]
\end{align}
independent of the planet's eccentricity $\sub{e}p$. This differs from the collision cross-section for a free-floating planet (equation~\ref{eq:sigma:collision:p} with $w=v_\infty$) by the additional term involving $\sub{v}c$. For Jupiter $\sub{v}c=13.1\,\kms$ and $\sub{v}{esc,p}=60.2\,\kms$, such that this makes no big difference, but does for Earth as $\sub{v}c=29.8\,\kms$ and $\sub{v}{esc,p}=11.2\,\kms$.

We also calculate the cross-section for collisions of ISOs with a moon, using the subscript `m' for properties of the moon and its orbit around the planet. After replacing $v_\infty\to w$, $\sub{R}p\to\sub{R}m$, $\sub{v}{esc,p}\to\sub{v}{esc,m}$, and $\sub{v}{c}^2\to\sub{v}{c,m}^2\equiv G(\sub{m}p+\sub{m}m)/\sub{a}m$, equation~\eqref{eq:sigma:collision} also gives the local cross-section for lunar collisions of ISOs entering the planet's Roche sphere with speed $w$. After averaging over the orientation of incoming ISOs and the planet orbit the cross-section for ISO collision with a moon is obtained as\footnote{We are not aware of previous publications of equations~\eqref{eq:sigma:collision} and~\eqref{eq:sigma:collision:moon}.}
\begin{align}
	\label{eq:sigma:collision:moon}
	\sub{\sigma}{coll,m} \doteq \pi\sub{R}m^2
		\left[1 + \frac{\sub{v}{esc,m}^2}{v_\infty^2} + \frac73\frac{\sub{v}{c,m}^2+\sub{v}c^2}{v_\infty^2} \right].
\end{align}
This result holds as long as $\sub{m}m\ll\sub{m}p\ll\sub{m}s$ and the lunar and planetary orbits are not in resonance.

The collision probability per unit surface area scales as $\sub{\sigma}{coll}/\pi R^2$, corresponding to the square brackets in equations~\eqref{eq:sigma:collision} and \eqref{eq:sigma:collision:moon}. For all moons in the Solar system, this quantity is smaller than for their host planets, because the large gulf between the respective escape speeds exceeds the additional contribution involving $\sub{v}{c,m}$ in equation~\eqref{eq:sigma:collision:moon}. See also \Sect{sec:coll+eject} for an application of equations~\eqref{eq:sigma:collision} and~\eqref{eq:sigma:collision:moon} to the Solar system.

%%%%%%%%%%%%%%%%%
%%%%%%%%%%%%%%%%%
%%%%%%%%%%%%%%%%%%%%%%%%%%%%%%%%%%%%%%%%%%%%%%%%%%%%%%%
\subsection{Capture by close encounters (slingshot)}
\label{sec:analytic:capture:strong}
We now use the impulse approximation to derive an analytic estimate for the capture cross-section in the limit of large $v_\infty$. This extends previous studies \citep[e.g.][]{Radzievskii1967, BandermannWolstencroft1970, RadzievskiiTomanov1977, PineaultDuquet1993} to planets on eccentric orbits. We also estimate the maximum capturable $v_\infty$ and the maximum binding energy of captured ISOs.

If the captured ISO has semi-major axis $a=GM/\sub{v}a^2$, energy conservation implies its barycentric speed just after the impulsive encounter with the planet to be
\begin{align}
	\label{eqs:v2:energy}
	v_2^2 = 2\sub{v}r^2 - \sub{v}a^2
\end{align}
and the energy change required for capture $\Delta E=-(v_\infty^2 + \sub{v}a^2)/2$.

The ISO trajectory in the planetocentric and barycentric frames is sketched in \autoref{fig:Orbit}. The incoming planetocentric asymptote of the ISO is characterised by the velocity $\vec{w}_1$ and offset $\sub{\vec{b}}p$, such that $\vec{r}(t) = \sub{\vec{b}}p + \vec{w}_1\,t$ at $t\to-\infty$. Since $\sub{\vec{b}}p\perp\vec{w}_1$, the directions $\uvec{w}_1$, $\sub{\uvec{b}}p$, and $\uvec{w}_1\cross\sub{\uvec{b}}p$ form a triad and we can express the planet's direction of motion uniquely as
\begin{align}
    \label{eq:vp:impact}
	\sub{\uvec{v}}p = 
	\cos\theta\,\uvec{w}_1+\sin\theta(\cos\sub{\psi}p\,\sub{\uvec{b}}p+\sin\sub{\psi}p\,\uvec{w}_1\cross\sub{\uvec{b}}p).
\end{align}
Here, $\theta\in[0,\pi]$ is the angle between $\sub{\vec{v}}p$ and $\vec{w}_1$ and $\sub{\psi}p\in[0,2\pi]$ the angle in the \emph{impact plane} (perpendicular to $\vec{w}_1$) between $\sub{\vec{b}}p$ and the projection of $\sub{\vec{v}}p$ onto the impact plane ($\sub{\psi}p$ is the tilt between the orbital plane and the plane spanned by $\uvec{w}_1$ and $\sub{\uvec{v}}p$). After the hyperbolic deflection by the angle $\sub{\theta}{def}$ the ISO's outgoing planetocentric velocity is (see also \autoref{fig:Orbit})
\begin{align}
    \label{eq:w:2:def}
    \vec{w}_2 = w\left[\cos\sub{\theta}{def}\uvec{w}_1-\sin\sub{\theta}{def}\sub{\uvec{b}}p\right]
\end{align}
and the ISO's barycentric speeds $v_{1,2}$ before and after encountering the planet are obtained by inserting equations~\eqref{eq:vp:impact} and~\eqref{eq:w:2:def} into~\eqref{eq:w:12}:%
\begin{subequations}
\label{eqs:v12:theta}
\begin{align}
	\label{eq:v:1:theta}
	v_1^2 &= \sub{v}p^2 + w^2 + 2\sub{v}pw\cos\theta,\\
	v_2^2 &= \sub{v}p^2 + w^2 + 2 \sub{v}pw(\cos\theta\cos\sub{\theta}{def}-\sin\theta\sin\sub{\theta}{def}\cos\sub{\psi}p).
\end{align}
\end{subequations}
Using the relations (see equations~\ref{eq:defl} and \ref{eq:e,a:b,w})
\begin{align}
	\label{eq:theta:def}
	\cos\sub{\theta}{def}
			= \frac{\sub{b}p^2w^4-G^2\sub{m}p^2}{\sub{b}p^2w^4+G^2\sub{m}p^2},
	\quad
	\sin\sub{\theta}{def}
			= \frac{2\sub{b}p w^2G\sub{m}p}{\sub{b}p^2w^4+G^2\sub{m}p^2},
\end{align}
the capture condition is obtained by inserting equations \eqref{eqs:v12:theta} and \eqref{eq:theta:def} into $v_1^2-v_2^2=v_\infty^2 + \sub{v}a^2$:
\begin{align}
	\label{eq:capt:condition:0}
	v_\infty^2 + \sub{v}a^2 = 4G\sub{m}p\sub{v}pw
	\frac{G\sub{m}p\cos\theta+\sub{b}pw^2\sin\theta\cos\sub{\psi}p}{G^2\sub{m}p^2 + \sub{b}p^2 w^4}.
\end{align}
The orientation $\theta$ and speed $w$ are related via equation~\eqref{eq:v:1:theta}, which together with~\eqref{eq:v:1} gives
\begin{align}
	\label{eq:C}
	2\sub{v}pw\cos\theta = C \equiv 2\sub{v}r^2 - \sub{v}p^2 - w^2 + v_\infty^2
	%\doteq \sub{v}c^2 - w^2 + v_\infty^2
	.
\end{align}
We may use this relation to eliminate $\theta$ from equation~\eqref{eq:capt:condition:0} and obtain the capture condition for $w$, $\sub{b}p$, and $\sub{\psi}p$ at given $v_\infty$ and $\sub{v}a$:
\begin{align}
	\label{eq:capt:condition:1}
	v_\infty^2 + \sub{v}a^2 = \frac{2G^2\sub{m}p^2}{G^2\sub{m}p^2+\sub{b}p^2w^4}
	\left[C+\cos\sub{\psi}p\frac{\sub{b}pw^2}{G\sub{m}p}
	\sqrt{4\sub{v}p^2w^2-C^2}\right].
\end{align}

%%%%%%%%%%%%%%%%%%%%%%%%%%%%%%%%%%%%%%%%%%%%%%%%%%%%%%%
\subsubsection{The capture cross section}
\label{sec:capture:impulse}
The calculation of the cross-section at infinity for capture is analogous to that for collision: after obtaining the local cross-section $\sub{\sigma}p$ the flux is averaged over all orientations and along the planet orbit. To find $\sub{\sigma}p$, we note that the solutions for $\sub{b}p$ and $\sub{\psi}p$ to equation~\eqref{eq:capt:condition:1} at fixed $w$, $\sub{v}p$, $v_\infty$, and $\sub{v}a$ lie on the circle 
\begin{align}
	\label{eq:capt:condition:2}
	(\sub{b}p\cos\sub{\psi}p-b_0)^2 + \sub{b}p^2\sin^2\!\sub{\psi}p = \rho^2
\end{align}
in the impact plane with offset and radius%
\begin{subequations}
\label{eqs:b0,rho}
\begin{align}
	\label{eq:b0}
	b_0 &= \frac{G\sub{m}p}{(v_\infty^2 + \sub{v}a^2)w^2}
	\sqrt{4w^2\sub{v}p^2-C^2},
	\\
	\label{eq:rho}
	\rho &= \frac{G\sub{m}p}{(v_\infty^2 + \sub{v}a^2)w^2}
	\sqrt{4w^2\sub{v}p^2-D^2}
\end{align}
\end{subequations}
with
\begin{align}
	\label{eq:D}
    D \equiv v_\infty^2+\sub{v}a^2-C
        = -2\sub{v}r^2 + \sub{v}p^2 + w^2 + \sub{v}a^2
\end{align}
(equivalent to relations given by \citealt{BandermannWolstencroft1970} and \citealt{RadzievskiiTomanov1977}). Since $\partial\rho/\partial\sub{v}a<0$, the cross-section of the planet to scatter an ISO onto a bound orbit with semi-major axis $0<a\le GM/\sub{v}a^2$ is $\sub{\sigma}p=\pi\rho^2$ and the local flux of ISOs scattered by the planet onto bound orbits is $F=n_1w\pi\rho^2$. Both depend (through $w$) on the orientation of the ISO orbit relative to the planet's direction of motion. Only those values of $w$ for which $b_0$ and $\rho$ remain real correspond to orientations that lead to capture, which implies $w_-\le w\le w_+$ with%
\begin{subequations}
\label{eq:w:limits}
\begin{align}
    \label{eq:w:min}
	w_- &= \max\left\{
		{\textstyle \sqrt{2\sub{v}r^2+v_\infty^2}}-\sub{v}p,\;
		\sub{v}p-{\textstyle \sqrt{2\sub{v}r^2-\sub{v}a^2}}\right\}, \\	
    \label{eq:w:max}
	w_+ &= \sub{v}p+{\textstyle\sqrt{2\sub{v}r^2-\sub{v}a^2}}.
\end{align}
\end{subequations}
To average the flux over all orientations $\alpha$, we use equation~\eqref{eq:dcosalpha} with these limits, which gives
\begin{align}
	\label{eq:F:alpha}
	\langle F\rangle_\alpha
	\doteq \frac{8\pi}{3}
		\frac{n_1\,G^2\sub{m}p^2\sub{v}p^2}{(v_\infty^2 + \sub{v}a^2)^2v_1}\,
		Y\left(\frac{1}{1+\sub{e}p\cos\sub{\eta}p},\frac{v_\infty^2}{\sub{v}p^2},\frac{\sub{v}a^2}{\sub{v}p^2}\right),
\end{align}
where
\begin{align}
	\label{eq:Y}
	Y(u,x,z) &\equiv 
		\left[\frac{3}{16y}\left(2u-1-z\right)^2
	  + \frac{3y}{8}\left(2u+1-z\right)
	  - \frac{y^3}{16}\right]_{y_-}^{y_+}
\end{align}
with $y_-\equiv \max\big\{\sqrt{2u+x}-1,1-\sqrt{2u-z}\big\}$ and $y_+\equiv1+\sqrt{2u-z}$.
%%%%%%%%%%%%%%%%%
\begin{figure}
	\includegraphics[width=\columnwidth]{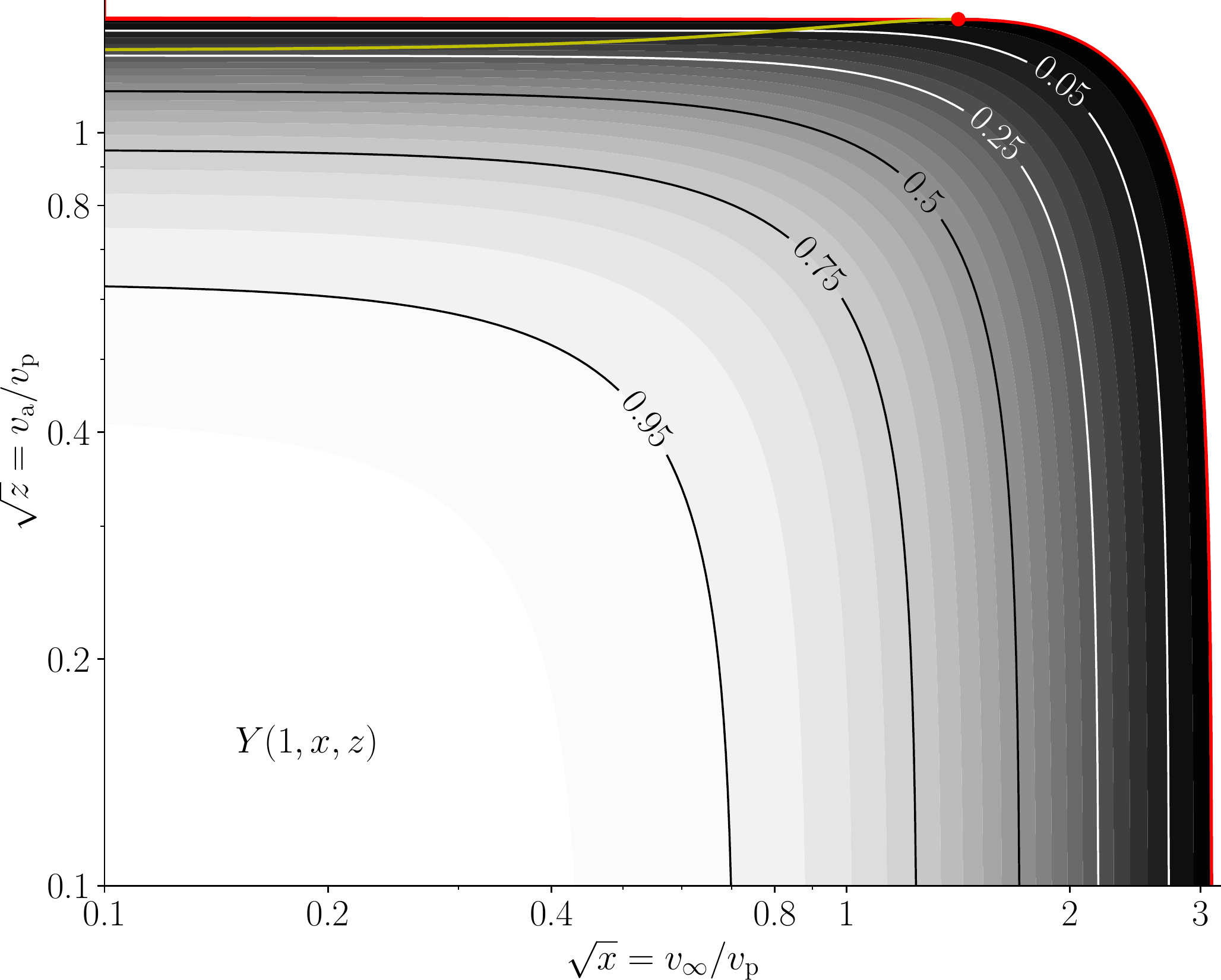}
	\caption{\label{fig:Y}
	Contours of the transfer function $Y(u=1,x,z)$ (defined in equation~\ref{eq:Y}) for capture of an ISO with asymptotic speed $v_\infty=\sqrt{x}\sub{v}p$ onto a bound orbit with energy $E\le-\frac12\sub{v}a^2=-\frac12z\sub{v}p^2$ via slingshot by a planet on a circular orbit with speed $\sub{v}p$. The red rim, beyond which $Y=0$, corresponds to the maximum possible $v_\infty$ at given $\sub{v}a$ and vice versa. To the right of the red point (at $v_\infty^2=4\sub{v}p^2-2\sub{v}r^2$ and $\sub{v}a^2=2\sub{v}r^2$) these maxima require a purely radial (reflective) orbit relative to the planet. The yellow curve is the border between the two values within the $\max\{\}$ operator in equation~\eqref{eq:w:min}: above that curve $\p Y/\p x=0$.
	}
\end{figure}
%%%%%%%%%%%%%%%%%
As can be seen in \autoref{fig:Y}, the \emph{transfer function} $Y$ essentially truncates the flux at the maximum possible $v_\infty$ and $\sub{v}a$, but for $v_\infty^2 + \sub{v}a^2\lesssim\sub{v}p^2$ deviates only little from $Y(u,0,0)=1$, see also \autoref{fig:Y}. Incidentally, the corresponding transfer function for the reverse process, ejection of a bound object, is generally larger than $Y$ (but of course has the same domain), see \App{app:eject}.

Finally, we must average the flux also over the eccentric planet orbit as in equation~\eqref{eq:F:ave:orbit} for the case of collisions. Unfortunately, owing to the complex dependence of $Y(u,x,z)$ on $\sub{\eta}p$ through all three arguments, the integral over $\sub{\eta}p$ does not result in a closed expression. For $v_\infty^2 + \sub{v}a^2\lesssim\sub{v}p^2$, however, $Y\sim1$ varies only weakly along the orbit (for mild eccentricities) and the behaviour of the flux is completely dominated by the remaining factors. For $v_\infty^2 + \sub{v}a^2\gtrsim\sub{v}p^2$, on the other hand, collisions with the planet, which have been neglected so far (but see below), become important rendering an exact average of the flux~\eqref{eq:F:alpha} inaccurate. Therefore, a reasonable approximation for weakly eccentric orbits and $v_\infty^2 + \sub{v}a^2\lesssim\sub{v}c^2$ is to exempt $Y$ from the orbit average and replace $\sub{v}p^2$ in the arguments to $Y$ by its orbit average. We have
\begin{align}
	\bigg\langle\frac{n_1\sub{v}p^2}{v_1}\bigg\rangle
	 = \frac{\sub{n}{iso}}{v_\infty}
	   \langle\sub{v}p^2\rangle
	 \doteq \frac{\sub{n}{iso}}{v_\infty}\sub{v}c^2,
\end{align}
independent of $\sub{e}p$. Dividing the average flux by $\sub{n}{iso} v_\infty$ finally obtains the cross-section for capture from close encounters
\begin{align}
	\label{eq:sigma:capture}
	\sigma(v_\infty|\sub{v}a)
	\approx \frac{8}{3}\pi\sub{a}p^2
		\frac{q^2\sub{v}c^6}{(v_\infty^2 + \sub{v}a^2)^2v_\infty^2}
		Y\left(1,\frac{v_\infty^2}{\sub{v}c^2},
			   \frac{\sub{v}a^2}{\sub{v}c^2}\right).
\end{align}
For $\sub{v}a\ll v_\infty=\sub{v}c$, the capture cross-section is roughly the area of a disc with radius $1.5q\sub{a}p$. For Jupiter ($\sub{v}c=13.1\,\kms$), this is 0.0075\,au, 16 times its radius, while the collision cross-section~\eqref{eq:sigma:collision} for this speed corresponds to a disc of 5 Jupiter radii. 
At fixed $\sub{v}a$ and $v_\infty$ (in the regime where the impulse approximation applies), capture cross-sections from planets within the same system scale roughly as $q^2/\sub{a}p$, which for Saturn, Uranus, and Neptune is, respectively, 0.048, 0.00056, and 0.00078 times that for Jupiter. Thus, except for a 5\% contribution from Saturn, slingshot captures into the Solar system are governed by Jupiter.

For circular binaries, a slightly wrong version of equation~\eqref{eq:sigma:capture} has been derived previously by \citeauthor{PineaultDuquet1993} (\citeyear{PineaultDuquet1993}, eqs.~11), who used an incorrect version of equation~\eqref{eq:w:min}, resulting in $Y<0$ above the yellow curve in \autoref{fig:Y}. In his investigation of the dynamical evolution of binaries in stellar clusters \cite{Heggie1975} considered the capture of a third body the ``most difficult to treat by any analytical means''. Nonetheless, integrating his equation 4.10 for the differential cross-section $\diff\sigma/\diff(\Delta E)$ obtains the form~\eqref{eq:sigma:capture} in the appropriate limit (as already noted by \citeauthor{PineaultDuquet1993}), except for the transfer function $Y$ and a factor of two.

%%%%%%%%%%%%%%%%%%%%%%%%%%%%%%%%%%%%%%%%%%%%%%%%%%%%%%%

%%%%%%%%%%%%%%%%%
\begin{figure}
	\includegraphics[width=\columnwidth]{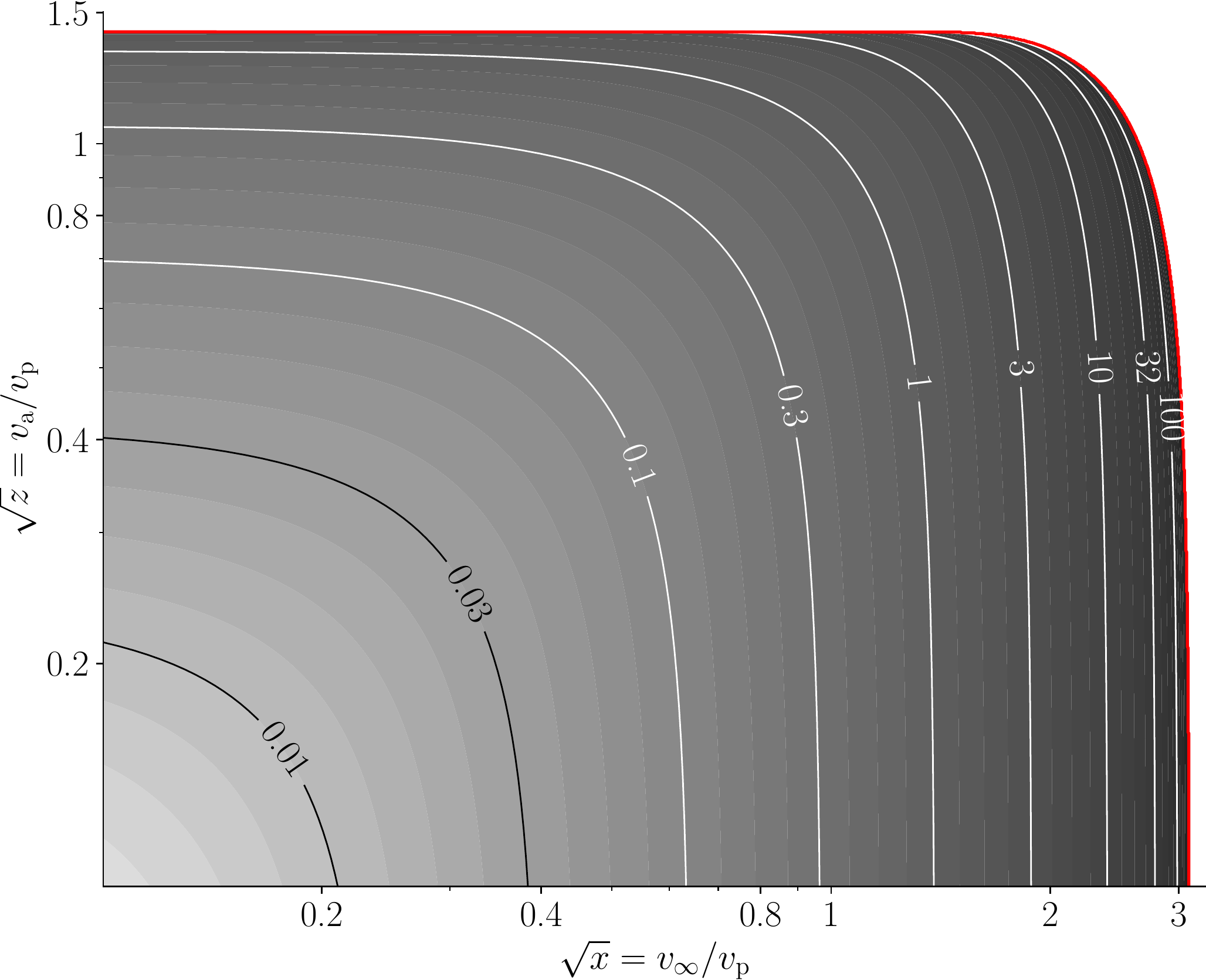}
	\caption{\label{fig:D}
	Contours of $\Theta_{\min}(u=1,x,z)$, defined in equation~\eqref{eq:Safronov:min}. Planets with radius $\sub{R}p$, mass $\sub{m}p$, and speed $\sub{v}p$ can only capture ISOs incoming with $v_\infty$ and onto orbits with $a<GM/\sub{v}a^2$ if $\Theta\equiv G\sub{m}p/\sub{R}p\sub{v}p^2 >\Theta_{\min}$. For trajectories not allowed by this criterion but within the red rim, the capture would require passage closer than the planets radius (see also text). The same condition is also required for the ejection of a bound object with semi-major axis $GM/\sub{v}a^2$ onto a hyperbolic orbit with asymptotic speed $v_\infty$ via a single slingshot (see \autoref{fig:Theta} for the values of $\Theta$ of known exoplanets). The red rim, corresponding to $\Theta_{\min}=\infty$, is identical to that in \autoref{fig:Y}.
	}
\end{figure}
%%%%%%%%%%%%%%%%%

\subsubsection{Collisions and the maximum $\sub{v}a$ and $v_\infty$}
\label{eq:analytic:strong:collide}
The maximum $v_\infty$ at given $\sub{v}a$ or the maximum $\sub{v}a$ at given $v_\infty$ are obtained by equating $w_-=w_+$, giving
\begin{subequations}%
\label{eq:ViVaMax}%
\begin{align}
	\label{eq:ViMax}
	v_{\infty,\max}^2
		&= 4\sub{v}p\left[{\textstyle\sqrt{2\sub{v}r^2-\sub{v}a^2}}+\sub{v}p^{}\right]
			- \sub{v}a^2,	\\
	\label{eq:VaMax}
	\sub{v}{a,\max}^2
		&= \begin{cases}
			4\sub{v}p\left[{\textstyle\sqrt{2\sub{v}r^2+v_\infty^2}}-\sub{v}p^{}\right]
			-v_\infty^2 & \text{for }v_\infty \ge \sqrt{2}\sub{v}c
			\\
			2\sub{v}r^2 & \text{otherwise},
			\end{cases}
\end{align}
\end{subequations}
corresponding to the red rim in \autoref{fig:Y}. For $v_\infty\ge\sqrt{2}\sub{v}c$ (right to the red dot in \autoref{fig:Y}), these extrema correspond to the situation where the incoming ISO moves parallel to the planet with impact parameter $\sub{b}p=0$: a purely radial orbit, which is `reflected' off the planet. In this case, the maximum capturable asymptotic speed is obtained for bare capture ($\sub{v}a=0$) when $\sub{v}{\infty,max}=2\sub{v}p(\sqrt{2}\sub{v}r/\sub{v}p+1)^{1/2}$, which for circular planet orbits is about $3.1\sub{v}c$, but larger for deflection off planets at perihel of an eccentric orbit. The maximum captured $\sub{v}a$ is obtained when $v_2=0$ (only possible at the apo-centre of a radial barycentric orbit) which gives $\sub{v}a=\sqrt{2}\sub{v}r$, corresponding in \autoref{fig:Y} to the red rim left of the red point.

In reality, of course, purely radial planetocentric orbits result in the ISO colliding with the planet rather being deflected by it. This implies that, even for circular planet orbits, the transfer factor $Y$ in the estimate~\eqref{eq:sigma:capture} is not quite correct, but over-predicts the capture cross-section and implies too large maximum capturable $v_\infty$ and captured $\sub{v}a$. In reality, a planet of a certain radius $\sub{R}p>0$ cannot effectuate captures in the same way as a point-like planet.

Taking the collision-condition into account when estimating the capture cross-section requires numerical treatment and results in estimates for collision-corrected transfer functions. This would need to be done for each value of the Safronov number
\begin{align}
	\label{eq:Safronov}
	\Theta \equiv \frac{G\sub{m}p}{\sub{R}p\sub{v}p^2}
		= \frac{\sub{v}{esc,p}^2}{2\sub{v}p^2}
		\doteq \frac{\sub{a}pq}{\sub{R}p},
\end{align}
which is a dimensionless measure of the (reciprocal) planet size.

We abstain from such an extensive treatment and instead consider the maximum capturable  and captured speeds, $v_\infty$ and $\sub{v}a$, possible for any given Safronov number $\Theta$. That is, we estimate where in \autoref{fig:Y} the red rim would be if collisions were taken into account. To this end we find, for every combination of $v_\infty$ and $\sub{v}a$, the largest $\sub{d}p$ of all possible capture orbits. At fixed $v_\infty$, $\sub{v}a$, and $w$, the largest $\sub{d}p$ is given by equation~\eqref{eq:dp} with impact parameter $\sub{b}p=b_0+\rho$. We then numerically find the maximum $\sub{d}p$ over all $w\in[w_-,w_+]$. Clearly, planets with radius $\sub{R}p$ larger than that maximum cannot effectuate the corresponding capture. Expressing this in terms of the Safronov number, capture requires that
\begin{align}
	\Theta > \Theta_{\min}\left(\frac{\sub{v}r^2}{\sub{v}p^2},
		\frac{v_\infty^2}{\sub{v}p^2},\frac{\sub{v}a^2}{\sub{v}p^2}\right),
\end{align}
with
\begin{align}
	\label{eq:Safronov:min}
	& \Theta_{\min}(u,x,z) \equiv \min_{y_-\le y\le y_+}\left\{y^{2}
	\left(\sqrt{B^2(u,x,y,z)+1}-1\right)^{-1}\right\},
	\\
	&B \equiv \frac{\sqrt{4y^2-(2u-1-y^2-z)^2}+\sqrt{4y^2-(2u-1-y^2+x)^2}}{x+z}.
\end{align}
\autoref{fig:D} shows the contours of $\Theta_{\min}(u=1,x^2,z^2)$ obtained by numerical optimisation, corresponding to the situation of a circular planet orbit. These contours are the extremal curves (like the red rim in \autoref{fig:Y}) for slingshots by planets with given Safronov number $\Theta$. Obviously from \autoref{fig:D}, the ability of a planet to capture an interstellar ISO, rather than collide with it, is diminished the smaller $\Theta$. First this affects only the maximum captureable speed $v_\infty$ (right edge in \autoref{fig:D}), but for $\Theta\lesssim2.5$, also the maximum $\sub{v}a$ is reduced.

For Jupiter, Saturn, Uranus, and Neptune, $\Theta\approx10.6$, 7.04, 4.93, and 9.43, respectively, all of which are well above unity, implying that captures by these planets are only mildly reduced by collisions. More detailed calculations for these planets, taking their eccentricities into account gives $v_{\infty,\max}=32.5$, 22.3\footnote{This contradicts the claim by \cite{Torbett1986} that Jupiter is the only planet in the Solar system able to capture an object with asymptotic speed $v_\infty=20\,\kms$. This suggests an error in Torbett's analysis, who used the same approximation but assumed a circular planet orbit, when we still find $v_{\infty,\max}=31.6\,\kms$ and $21.6\,\kms$ for Jupiter and Saturn, respectively.}, 14.5, and $13.0\,\kms$.

%%%%%%%%%%%%%%%%%%%%%%%%%%%%%%%%%%%%%%%%%%%%%%%%%%%%%%%
\subsubsection{Limitation of the impulse approximation}
\label{sec:analytic:strong:limit}
We already discussed the limitations of validity of the capture cross-section~\eqref{eq:sigma:capture} at high $v_\infty^2 + \sub{v}a^2$ owed to the neglect of collisions. We now consider the situation at low $v_\infty^2 + \sub{v}a^2$, i.e.\ small changes in the orbital energy. Such small changes can be effectuated already in rather wide encounters when the approximation of the planetary influence as an impulsive change of the barycentric orbit is clearly incorrect.

The assumptions underpinning the impulse approximation are only valid as long as most of the deflection occurs within the planet's Roche sphere with radius $\sub{R}{Roche}=q^{1/3}\sub{a}p$. We may estimate the size of the deflection region by the semi-latus rectum $\sub{s}p=\sub{b}p^2w^2/G\sub{m}p$ of the planetocentric ISO orbit. From equations~\eqref{eqs:b0,rho} we find for $v_\infty^2 + \sub{v}a^2\ll\sub{v}p^2$ that $\sub{b}p\sim2G\sub{m}p\sub{v}p/(v_\infty^2 + \sub{v}a^2)w$ and hence $\sub{s}p\sim4G\sub{m}p\sub{v}p^2/(v_\infty^2 + \sub{v}a^2)^2$, such that $\sub{s}p\lesssim\sub{R}{Roche}$ implies
\begin{align}
	\label{eq:impulse:condition:1}
	- 2\Delta E = 
	v_\infty^2 + \sub{v}a^2 \gtrsim
		2\,q^{1/3} \sub{v}p^2.
\end{align}
However, our numerical results in the next section indicate that equation~\eqref{eq:sigma:capture} remains valid down to much smaller values for $|\Delta E|$, namely for
\begin{align}
	\label{eq:impulse:condition:numerical}
	- 2\Delta E = v_\infty^2 + \sub{v}a^2 \gtrsim 3q\,\sub{v}p^2.
\end{align}
This is astonishing, since even allowing $\sub{s}p\lesssim\sub{a}p$, i.e.\ the 
deflection to occur over a region as large as the planet orbit, one only finds
$v_\infty^2 + \sub{v}a^2\gtrsim2\,q^{1/2}\sub{v}p^2$, still much larger than~\eqref{eq:impulse:condition:numerical}. Alternatively, if one demands that the capture cross-section~\eqref{eq:sigma:capture} is smaller than that for crossing the planet's Roche sphere (estimated via equation~\ref{eq:sigma:collision} with $\sub{R}{p}$ replaced by $\sub{R}{Roche}$), one finds $v_\infty^2 + \sub{v}a^2\gtrsim q^{2/3} \sub{v}p^2$ still larger than the numerical result~\eqref{eq:impulse:condition:numerical}.

%%%%%%%%%%%%%%%%%%%%%%%%%%%%%%%%%%%%%%%%%%%%%%%%%%%%%%%
\subsection{Capture by wide encounters}
\label{sec:analytic:capture:weak}
Small changes $\Delta E= -\tfrac12(v_\infty^2 + \sub{v}a^2)$ to the orbital energy of the ISO can already be effectuated by weak interactions, when the actual ISO orbit deviates only little from that in the absence of the planet. The natural tool for estimating the energy change during such a wide encounter is perturbation theory. In barycentric coordinates the Hamiltonian can be split as $H(\vec{x},\vec{v},t)=H_0(\vec{x},\vec{v})+H_1(\vec{x},t)$ with $H_0=\tfrac12\vec{v}^2-GM/|\vec{x}|$, when to first-order the energy change is
\begin{align}
	\label{eq:dE:pert:1}
	\Delta E \approx \int \left(\pdiff{H_1}{t}\right)_{\vec{x}_0(t)} \diff t,
\end{align}
i.e.\ $\partial H_1/\partial t$ is evaluated along the unperturbed barycentric hyperbola instead of the true trajectory. The integral~\eqref{eq:dE:pert:1} is not expressible in closed form and requires numerical treatment \citep[see][for a similar approach]{BreakwellPerko1974}. We may simplify the problem by assuming a circular planet orbit and replacing $H_1$ with its quadrupole approximation \citep[e.g.][equation A9]{AlyEtAl2015}
\begin{align}
	\pdiff{H_1}{t} &\approx -\frac{3}{2}\frac{GM}{r^3}\sub{a}p^2\,q\,\sub{\Omega}p
		\sin^2\vartheta\,	
		\sin2(\varphi-\phi_0-\sub{\Omega}pt),
\end{align}
where $\phi_0$ is the planetary azimuth at ISO periapse and $(r,\vartheta,\varphi)$ are the polar coordinates of the ISO along its orbit. At $v_\infty\to0$, we may approximate the ISO orbit as parabolic (see Appendix~\ref{app:para}), when for the most favourable situation of a co-planar ISO orbit ($\sin\vartheta=1$) we obtain (using equations~\ref{eq:para:r:phi}-\ref{eq:para:t:phi})
\begin{align}
	\label{eq:dE:quadrupole}
	\Delta E &\simeq \frac12 \frac{\sub{a}p^{3/2}}{s^{3/2}}
		q\sub{v}c^2 \sin2\phi_0 \,I\left(\frac{s^{3/2}}{\sub{a}p^{3/2}}\right)
\end{align}
with $s=b^2v_\infty^2/GM$ the ISO's semi-latus rectum. Here,
\begin{align}
	\label{eq:I(x)}
	I(x) &\equiv 12\int_{-\pi/2}^{\pi/2}\cos^2\!u\,\cos\left(4u-x\big[\tan u + \tfrac13 \tan^3\!u \big] \right)\diff u,
\end{align}
which is always positive, but decays very quickly with increasing $|x|$, which is essentially the ratio $|\sub{\Omega}p/\Omega_\varphi|$ between the azimuthal frequencies of planet and ISO. The larger this ratio, the more strongly the integrand in~\eqref{eq:I(x)} oscillates, when negative and positive contributions mostly cancel. The maximum negative energy change occurs for planetary phases $\phi_0=-\tfrac14\pi$ and $\tfrac34\pi$, but beyond the quadrupole approximation the former is favourable, which corresponds to passing the star on the same side as the planet. For retrograde encounters, $\Delta E$ is given by the same expression with $I(x)$ replaced with $I(-x)$. This is also positive but much smaller than $I(x)$, such that capture by wide interactions occurs preferentially from prograde orbits. % For inclined ISO orbits, the azimuthal frequency drops, approaching zero for $i=90^\circ$, and so does $\Delta E$.

%%%%%%%%%%%%%%%%%
\begin{figure}
	\includegraphics[width=\columnwidth]{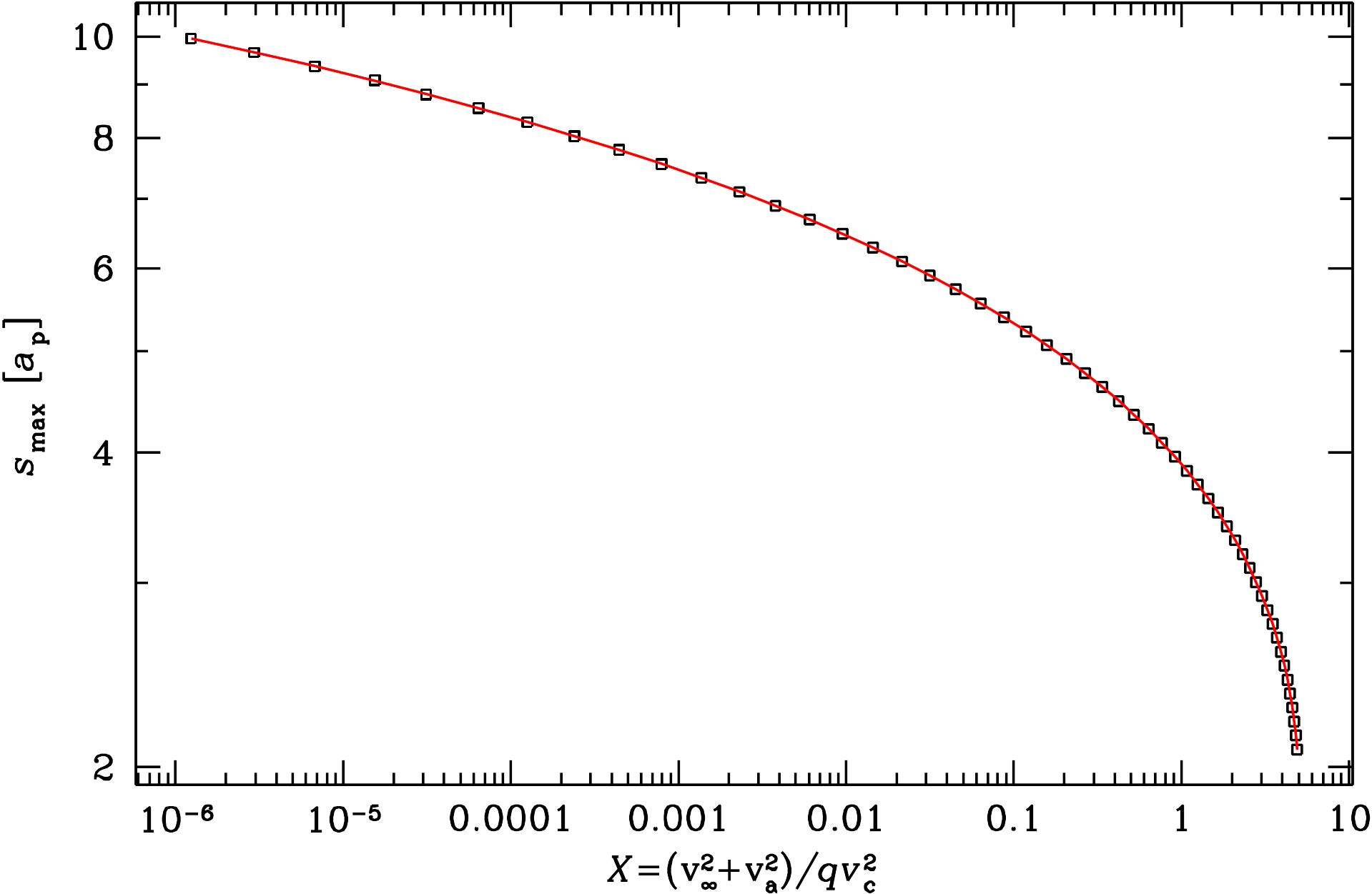}
	\caption{\label{fig:Smax:X}
	The maximum semi-latus rectum $s=b^2v_\infty^2/GM$ of incoming ISO orbits for which the energy change can be as large as $\Delta E=-\frac12 qX\sub{v}c^2$. This maximum is obtained for a prograde co-planar orbit leading the planet (assumed on a circular orbit) by $-\phi_0=\pi/4$ at peri-centre. Points: relation from numerically obtained $\Delta E$ (see text), curve: parametric fit with asymptote $\sub{s}{max}\propto|\ln X|^p$ where $p\sim\tfrac23$ at small $X$.
	}
\end{figure}
%%%%%%%%%%%%%%%%%

Defining 
\begin{align}
    \label{eq:J}
    J(x) \equiv x^{-3/2}\,I(x^{3/2})
\end{align}
and inverting equation~\eqref{eq:dE:quadrupole} for $\sin2\phi_0=-1$, obtains the maximum semi-latus rectum for capture 
\begin{align}
	\label{eq:s:X}
	s_{\max} = \sub{a}p J^{-1}(X),
\end{align}
with the dimensionless measure
\begin{align}
	\label{eq:X}
	X \equiv \frac{1}{q} \frac{v_\infty^2 + \sub{v}a^2}{\sub{v}c^2} = -\frac{2}{q} \frac{\Delta E}{\sub{v}c^2}
\end{align}
for the energy change. \autoref{fig:Smax:X} plots the relation~\eqref{eq:s:X} obtained in this way and by numerical quadrature of~\eqref{eq:I(x)} as points. Obviously, the semi-latus rectum (and hence peri-centre) of incoming ISO orbits that can just be captured increases only very slowly towards vanishing $\Delta E$, i.e.\ $X\to0$: the red curve in \autoref{fig:Smax:X} asymptotes to $\sub{s}{max}\propto|\ln X|^p$ with $p\sim\tfrac23$ in that limit.

The contribution to the capture cross-section in the limit $v_\infty\to0$ is $\sigma=\pi b^2 =\pi \sub{a}p s \sub{v}c^2/v_\infty^2$. Hence, 
\begin{align}
	\label{eq:sigma:wide}
	\sigma = \pi \sub{a}p \frac{\sub{v}c^2}{v_\infty^2}
    \int \tdiff{p(s|X)}{s} s\,\diff s,
\end{align}
where $\diff p(s|X)$ is the probability that an ISO orbit with semi-latus rectum $\in[s,s+\diff s]$ suffers an energy change $\Delta E\le -\frac12\sub{v}c^2qX$ (note that $\int \diff p<1$). From simple scaling relations we expect that $\diff p/\diff s$ depends only on the ratio $s/s_{\max}(X)$, i.e.\ is of the form
\begin{align}
	\label{eq:p(s|X)}
    \tdiff{p(s|X)}{s} = \frac1{\sub{a}p} P_{\!s}\left(\frac{s}{s_{\max}(X)}\right)
\end{align}
with some function $P_{\!s}(\xi)$ which accounts for the averaging over ISO orientations and planetary phase. While $P_{\!s}(\xi)$ cannot be worked out analytically, it must be a decreasing function and vanish for $\xi>1$. Inserting~\eqref{eq:p(s|X)} and~\eqref{eq:s:X}  into~\eqref{eq:sigma:wide} obtains
\begin{align}
	\label{eq:sigma:capture:wide}
	\sigma(v_\infty|\sub{v}a) &= \pi \sub{a}p^2 \frac{\sub{v}c^2}{v_\infty^2}\,\langle \xi\rangle_P\,J^{-1}(X),
\end{align}
where our ignorance of the function $P(\xi)$ has been reduced to the single factor 
\begin{align}
    \label{eq:<xi>} \textstyle
    \langle \xi\rangle_P \equiv \int P_{\!s}(\xi) \xi\diff\xi.
\end{align}
The numerical experiments in the next section confirm this approximation and suggest that $\langle\xi\rangle_P\approx\frac1{12}$. Other than for close encounters, where according to equation~\eqref{eq:sigma:capture} $\sigma$ scales like $X^{-2}$, the dependence of $\sigma$ on $X$ for wide encounters is rather weak, increasing shallower than a power law towards $X\to0$.

%%%%%%%%%%%%%%%%%%%%%%%%%%%%%%%%%%%%%%%%%%%%%%%%%%%%%%%
\subsection{Capture by the star?}
\label{sec:capture:star}
Since the star is not at rest but also moves in the barycentric frame, it can in principle also effectuate capture. We now assess the importance of this possibility. First, we note that captures by wide interactions are always dominated by the planet, simply because the contributions of planet and star to the binary quadrupole moment scale as $q$ and $q^2$, respectively.

For close interactions with the star, the impulse approximation used in our treatment of close interactions with the planet is not really appropriate. Howevever, we may still estimate an upper limit for the maximum energy change by assuming that the star moves with constant barycentric velocity and speed $\sub{v}s\doteq q\sub{v}c$. Then the maximum energy change occurs if the ISO passes just in front of the star on a grazing orbit, when
\begin{align}
    |\Delta E|\lesssim  \frac{2 q v_\infty \sub{v}{c}} {1+2v_\infty^2/\sub{v}{esc,s}^2} < 2 q v_\infty \sub{v}{c}
\end{align}
where $\sub{v}{esc,s}^2=2G\sub{m}s/\sub{R}s$ is the escape speed from the stellar surface. Thus, the maximum $v_\infty$ that can just be captured is $v_{\infty,\max}\simeq 4q\sub{v}c$, which is smaller than for capture by the planet by a factor $\sim q$. Thus, capture by the star is insignificant compared to that by the planet.
 
%%%%%%%%%%%%%%%%%%%%%%%%%%%%%%%%%%%%%%%%%%%%%%%%%%%%%%%
\section{Simulations of ISO capture}
\label{sec:sigma:sim}
We now numerically calculate the cross-section for capturing an ISO and compare it to the analytical estimates \eqref{eq:sigma:capture} and \eqref{eq:sigma:capture:wide}. Following \cite{ValtonenInnanen1982}, we integrate many orbits from random incoming trajectories, i.e.\ we randomly throw test particles at the planet-star binary and record those that `stick', but do not restrict ourselves to circular planet orbits. We use barycentric coordinates with the planet orbit in the $x$-$y$ plane, when the asymptotic impacting ISO orbit at $t\to-\infty$ is specified as
\begin{align}
	\label{eq:comet:asymp}
	\vec{r}(t) = \vec{b} + t \vec{v}_\infty
\end{align}
with vectors $\vec{b}\perp\vec{v}_\infty$, which are parameterised as
\begin{align}
	\vec{b} = b
	\begin{pmatrix}
		-\sin\phi\cos\psi - \cos\phi\sin\beta\sin\psi \\
		\phantom{-}\sin\phi\cos\psi - \sin\phi\sin\beta\sin\psi \\
		\phantom{- \sin\phi\cos\psi + \cos\phi}\cos\beta\sin\psi
	\end{pmatrix},
	\;
	\vec{v}_\infty = v_\infty
	\begin{pmatrix}
		\cos\phi\cos\beta \\
		\sin\phi\cos\beta \\
		\phantom{\cos\phi}\sin\beta
	\end{pmatrix}.
\end{align}
Here, $\beta$ is the ecliptic latitude and $\phi$ the azimuth of the direction $\uvec{v}_\infty$, while $\psi$ is the angle in the impact plane perpendicular to $\uvec{v}_\infty$. The incoming orbit~\eqref{eq:comet:asymp} is the asymptote of a unique hyperbolic orbit around the binary's barycentre, hereafter denoted the \emph{incoming hyperbola}. We choose $t=0$ to refer to the periapse of this hyperbola. With these specifications the capture cross section is
\begin{align}
	\label{eq:sigma:num}
	\sigma(v_\infty|\sub{v}a^2) &= \int_0^\infty b\diff b
		 	\int_0^{2\pi} \diff\psi 
			\int_{-\pi/2}^{\pi/2} \frac{\cos\beta\,\diff\beta}{2}
		 	\int_0^{2\pi} \frac{\diff\phi}{2\pi}
		 	\quad\times \nonumber \\
			&\phantom{=}
		 	\int_0^{2\pi} \frac{\diff\sub{\ell}p}{2\pi}\;
			H\big(-\tfrac12\sub{v}a^2-\sub{E}{final}\big),
\end{align}
where $\sub{\ell}p$ is the orbital phase (mean anomaly) of the planet at $t=0$\footnote{The planet's argument of periapse $\sub{\omega}p$ is redundant with the azimuth $\phi$ of the incoming orbit and we can set $\sub{\omega}p=0$ without loss of generality. Similarly, for circular planet orbits $\sub{\omega}p=0$ and $\sub{\ell}p=0$, without loss of generality.}.
$H$ denotes the Heaviside function and the final energy $\sub{E}{final}$ is a (non-trivial) function of $\vec{b}$, $\vec{v}_\infty$, and $\sub{\ell}p$ obtained via numerical orbit integration.

%%%%%%%%%%%%%%%%%%%%%%%%%%%%%%%%%%%%%%%%%%%%%%%%%%%%%%%
\subsection{Numerical method}
\label{sec:sigma:sim:numerics}
We now describe the numerical method to estimate the cross section in sufficient detail for anybody to reproduce it. Readers who are merely interested in the results may skip this sub-section.

%%%%%%%%%%%%%%%%%%%%%%%%%%%%%%%%%%%%%%%%%%%%%%%%%%%%%%%
\subsubsection{Initial conditions}
\label{sec:sigma:sim:IC}
We evaluate the integral~\eqref{eq:sigma:num} via Monte-Carlo integration with the main difference to \citeauthor{ValtonenInnanen1982} that we use many more individual orbits, enough to have at least $10^3$ capture events for each speed $v_\infty$ considered. The Monte-Carlo estimate for the capture cross-section is simply
\begin{align}
	\sigma = \pi b_{\max}^2 \frac{\sub{N}{captured}\pm\sqrt{\sub{N}{captured}}}{\sub{N}{sampled}},
\end{align}
where $b_{\max}$ is the maximum impact parameter $b$ sampled. We use
\begin{align}
	\label{eq:bmax}
	b_{\max} = p_{\max} \left[1 + \frac{2GM}{p_{\max}^{}v_\infty^2}\right]^{1/2},
\end{align}
such that the incoming hyperbola's periapse radius is at most
\begin{align}
	\label{eq:qmax}
	p_{\max} = \sub{a}p+ \sub{d}{p,max}+ \sub{R}{Roche}.
\end{align}
With this choice all incoming trajectories that pass the planet at distances $\sub{d}p\lesssim\sub{d}{p,max}$ are sampled (the addition of $\sub{R}{Roche}$ accounts for deviations of the actual orbits from simple hyperbolae). The correct choice of the parameter $\sub{d}{p,max}$ is critical for the validity of our method as detailed in section~\ref{sec:sigma:sim:param}.

Having drawn appropriate $\vec{b}$ and $\vec{v}_\infty$, the actual initial position and velocity for the numerical orbit integration are taken to be those of the incoming hyperbola when it first reaches the starting radius $r_0\gg\sub{a}p$ (the choice of $r_0$ is discussed in section \ref{sec:sigma:sim:param} below). This initial condition corresponds to a time $t=-T$, obtained from equation~\eqref{time:hyper:eta}, which we use to set the planet initial mean anomaly to $\sub{\ell}p=-\sub{\Omega}p T$ with $\sub{\Omega}p^2=GM/\sub{a}p^3$ the orbital frequency of the planet.

%%%%%%%%%%%%%%%%%%%%%%%%%%%%%%%%%%%%%%%%%%%%%%%%%%%%%%%
\subsubsection{Orbit integration}
\label{sec:sigma:sim:orbit}
The orbits are numerically integrated with an integrator that constructs the trajectory from alternating Kepler orbits around star and planet, and hence treats close encounters correctly, see Appendix~\ref{app:integrator} for details. We set the accuracy parameter (see equation~\ref{eq:T}) to $2\times10^{-6}$, when a few hundred steps per orbit are required, and in the case of a circular planet orbit the local and global errors of the Jacobi integral $J=E-\sub{\Omega}pL_z$ are $10^{-9}$ and $10^{-8}$ times $GM/\sub{a}p$, respectively. Of course, more accurate integration is possible at the expense of higher computational costs, but has no benefit since our uncertainties are dominated by the shot noise of the Monte-Carlo approach.

%%%%%%%%%%%%%%%%%
\begin{figure*}
	\includegraphics[width=120mm]{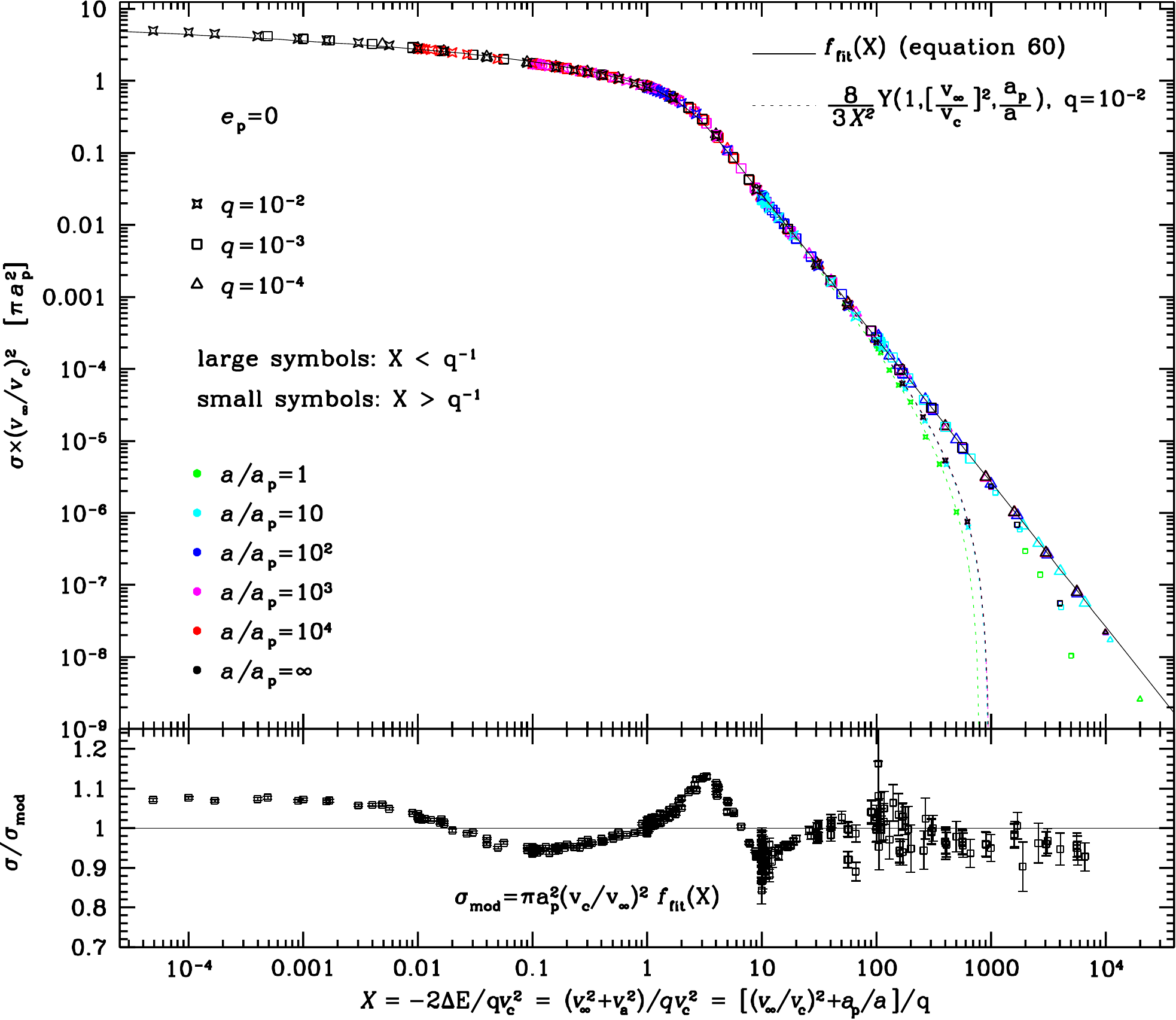}
	\caption{\label{fig:Sigma:X}
	\textbf{Top:} the numerically calculated cross-section $\sigma$, scaled by $(v_\infty/\sub{v}c)^2$, for capture by a planet on a circular orbit ($\sub{e}p=0$) with semi-major axis $\sub{a}p$ and radius $\sub{R}p=10^{-3}\sub{a}p$ is plotted against the dimensionless measure $X$ (equation~\ref{eq:X}) of the minimum energy change for various values of the planet-to-star mass ratio $q$ and the maximum semi-major axis $a$ for the initial captive orbit. Only values for $a$ with at least 100 simulated captures are included, such that the statistical error of the Monte-Carlo estimate for $\sigma$ is at most 10\% (smaller than the symbol sizes), but often much less. The thin full curve is a fitting function (equation~\ref{eq:f(X):fit}), which approaches the theoretical asymptotes at small and large $X$. The dotted lines are predictions for $q=10^{-2}$ including the transfer factor $Y$. \textbf{Bottom:} Ratio of the numerically calculated $\sigma$ to $\pi\sub{a}p^2(\sub{v}c/v_\infty)^2\sub{f}{fit}(X)$ for all data at $X<q^{-1}$.
	}
\end{figure*}
%%%%%%%%%%%%%%%%%
We integrate each orbit either until time $t=T$ (when the incoming hyperbola reaches the starting radius $r_0$ again on its way back out), or until the orbit passes an apo-apse (a sign change  of the radial velocity from $+$ to $-$) outside of the planet's Roche sphere, or until a collision with star or planet, whatever occurs first. For each orbit we record the initial parameters, the total and rms local errors of the Jacobi integral (for circular planet orbits only), the times and distances of closest approaches to star and planet, and the barycentric osculating orbital elements at the end of integration.

%%%%%%%%%%%%%%%%%%%%%%%%%%%%%%%%%%%%%%%%%%%%%%%%%%%%%%%
\subsubsection{Avoiding unnecessary integrations}
The vast majority of initial conditions generated in our Monte-Carlo approach pass through the binary like 'Oumuamua through the Solar system and do not result in capture, because they do not come close enough to the planet. Numerical integration of such interlopers is best avoided when calculating the capture cross section. Close encounters with the planet can be constructed by integrating initial conditions close to the planet both backward and forward in time \citep{SirajLoeb2019}. However, the probability for such events to arise from randomly incoming orbits cannot be obtained rigorously (only within the impulse approximation) and, hence, neither can be the capture cross-section with this method.

Instead, we only numerically integrate orbits for initial conditions whose incoming hyperbola actually comes closer to the planet than $\sub{d}{p,max}$. This is a much more stringent condition than that it comes closer than $\sub{a}p+\sub{d}{p,max}$ to the barycentre (satisfied by all initial conditions), especially for $\sub{d}{p,max}<\sub{a}p$. Calculating the closest approach $\sub{d}{p,hyp}$ of the incoming hyperbola to the planet requires careful numerical minimisation (for some orbits there are two minima) but is still $\sim10^{\text{2-3}}$ times faster than a full orbit integration, substantially cutting the overall computational costs, in particular for large $v_\infty$, when capture becomes ever less likely and requires very close encounters.

%%%%%%%%%%%%%%%%%
\begin{figure*}
	\includegraphics[width=120mm]{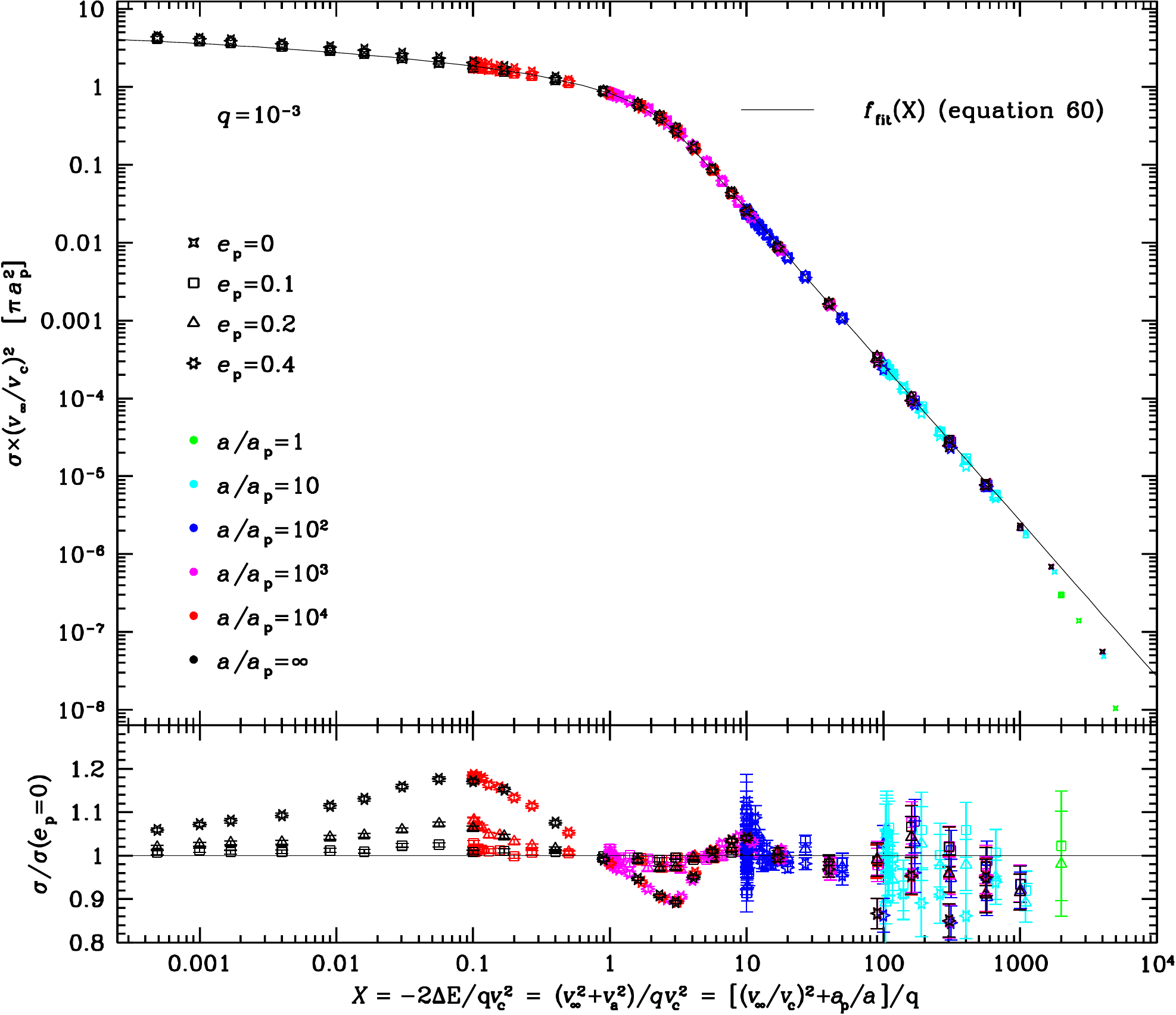}
	\caption{\label{fig:Sigma:X:ecc}
	\textbf{Top}: same as \autoref{fig:Sigma:X} but for different planet eccentricities $\sub{e}p$ (as indicated) and at fixed mass ratio $q=10^{-3}$. \textbf{Bottom}: ratio of the capture cross section at $\sub{e}p>0$ to that obtained for $\sub{e}p=0$ for the same $v_\infty,\,\sub{v}a$ (or $a$) plotted against the dimensionless measure $X$ for the minimum energy change suffered by the captured ISO.
	}
\end{figure*}
%%%%%%%%%%%%%%%%%

%%%%%%%%%%%%%%%%%%%%%%%%%%%%%%%%%%%%%%%%%%%%%%%%%%%%%%%
\subsubsection{Tuning the numerical parameters}
\label{sec:sigma:sim:param}
Apart from the accuracy parameter of the orbit integrator, our method has two more numerical parameters introduced in section~\ref{sec:sigma:sim:IC}: the initial radius $r_0$ of the numerical orbit integrations and the parameter $\sub{d}{p,max}$.

%%%%%%%%%%%%%%%%%%%%%%%%%%%%%%%%%%%%%%%%%%%%%%%
\paragraph*{\boldmath The choice of the starting radius $r_0$.}
%\label{sec:param:R}
Starting the numerical integration not at infinity but at $r_0$ incurs an error that to good approximation is given by the effect of the binary's quadrupole integrated from infinity to $r_0$. The amplitude of the quadrupole of the binary's orbit-averaged potential at barycentric radius $r$,% given in equation~\eqref{eq:Phi2},
\begin{align}
	\label{eq:Phi2}
	\Phi_2(r) = \sqrt{\frac38} \frac{\sub{a}p^2q}{(1+q)^2} \frac{GM}{r^3}
\end{align}
\citep[e.g.][equation A9]{AlyEtAl2015}, is therefore an estimate for the energy error contracted by starting the integration at $r$. Demanding that this error is much smaller than $|\Delta E|$ required for capture gives the condition
\begin{align}
	r_0^3 \gg \sqrt{\frac32}
		\frac{q\sub{a}p^2}{(1+q)^2}\frac{GM}{v_\infty^2 + \sub{v}a^2}
		\approx \frac{G\sub{m}p\sub{a}p^2}{v_\infty^2 + \sub{v}a^2}.
\end{align}
In practice we set 
\begin{align}
	r_0 = 10 \left(\frac{G\sub{m}p\sub{a}p^2}{v_\infty^2}\right)^{1/3},
\end{align}
or $r_0=20\sub{a}p$, whichever is larger.

%%%%%%%%%%%%%%%%%%%%%%%%%%%%%%%%%%%%%%%%%%%%%%%
\paragraph*{\boldmath The choice of $\sub{d}{p,max}$.}
%\label{sec:param:dmax}
The choice of $\sub{d}{p,max}$ is critical for the validity of our method: if too small some capture events will be missed because the orbit is not integrated, and if too large many non-capture events will be integrated unnecessarily. Moreover, the most suitable value depends strongly on $\vec{v}_\infty$. In practice, we leave a 10\% safety margin of values for $\sub{d}p$ which are integrated but never result in capture by ensuring that
\begin{align}
	\label{eq:crit:dmax}
	\max_{\mathrm{captures}}
	%\sub{\max}{bare\;capture}
	\{\sub{d}{p},\,\sub{d}{p,hyp}\} < 0.9\, \sub{d}{p,max},
\end{align}
where $\sub{d}{p}$ and $\sub{d}{p,hyp}$ denote closest-approach distances to the planet of the actual ISO orbit and the incoming hyperbola, respectively, while the maximum is over all initial conditions that have been integrated (i.e.\ for which $\sub{d}{p,hyp}<\sub{d}{p,max}$) and resulted in capture. In order to enforce this criterion, if $\sub{d}{p,max}$ is ever found to violate~\eqref{eq:crit:dmax} by a newly integrated orbit, it is increased appropriately (this does not require a complete re-start of the Monte-Carlo sampling as long as care is taken to adapt the previous sampling to the new value). This method works extremely well if started with slightly too small initial $\sub{d}{p,max}$.

%%%%%%%%%%%%%%%%%%%%%%%%%%%%%%%%%%%%%%%%%%%%%%%%%%%%%%%
\subsection{Results}
\label{sec:sigma:sim:results}
We use these methods to simulate, for $q=10^{-2},\,10^{-3},\,10^{-4}$ and a grid of values for $v_\infty$, as many ISO orbits as necessary, but no fewer than $10^7$, to generate at least $10^3$ bare captures, unless the number of captures exceeds $10^5$ (when we terminate the runs before $10^7$ orbits are reached). Here, `bare capture' refers to any bound ISO orbit, regardless of its semi-major axis, i.e.\ $\sub{E}{final}<0$. We assume that planet and star have radii $\sub{R}p=10^{-4}\sub{a}p$ and $\sub{R}s=10^{-3}\sub{a}p$, when the case $q=10^{-3}$ approximately corresponds to the Sun-Jupiter system.

%%%%%%%%%%%%%%%%%
\begin{figure*}
	\includegraphics[width=178.5mm]{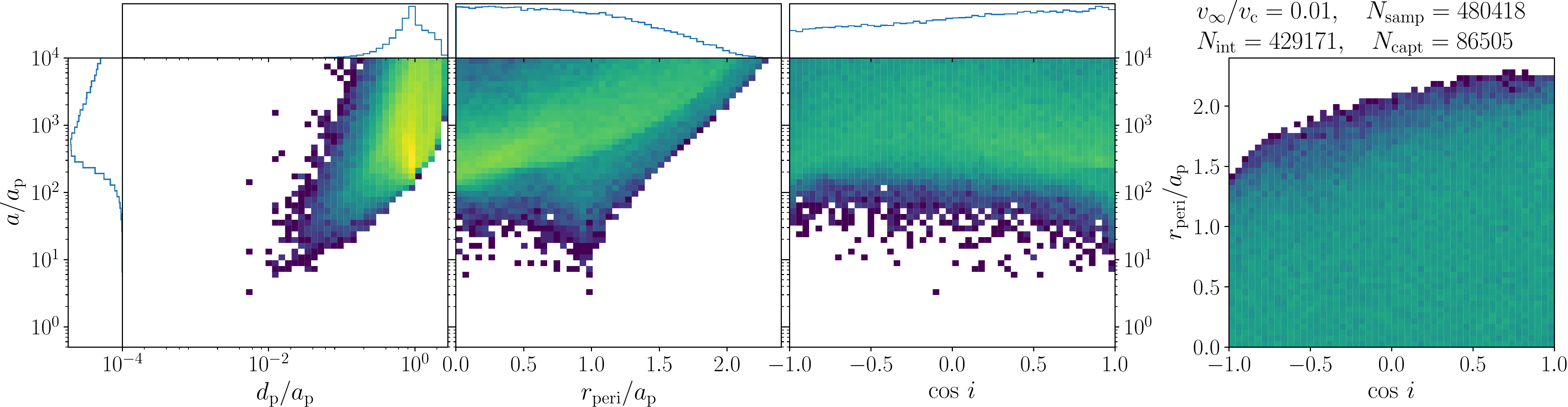}
	\vspace{1mm}
    \includegraphics[width=178.5mm]{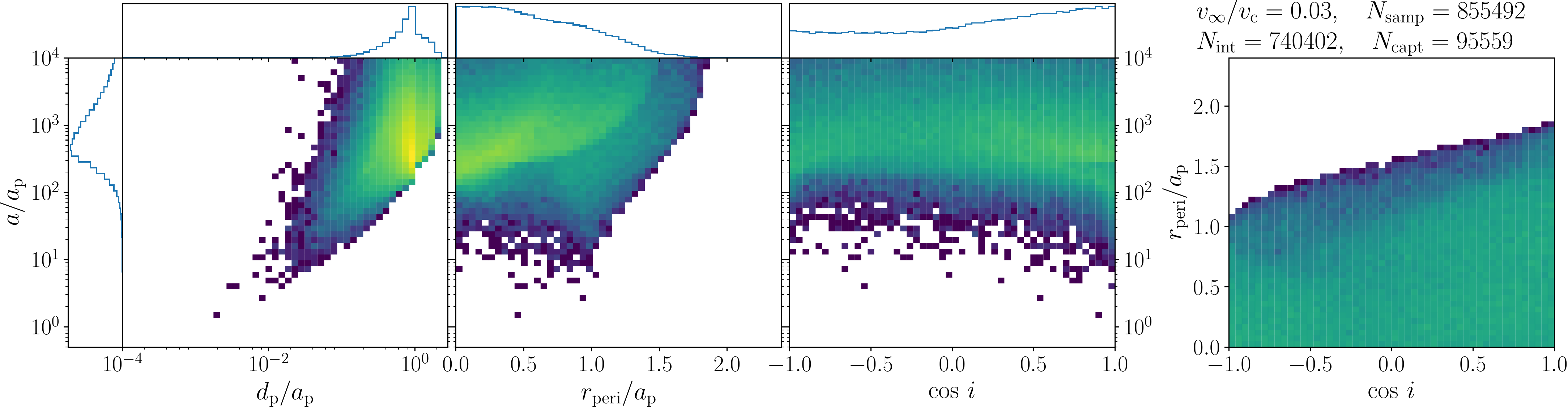}
	\vspace{1mm}
	\includegraphics[width=178.5mm]{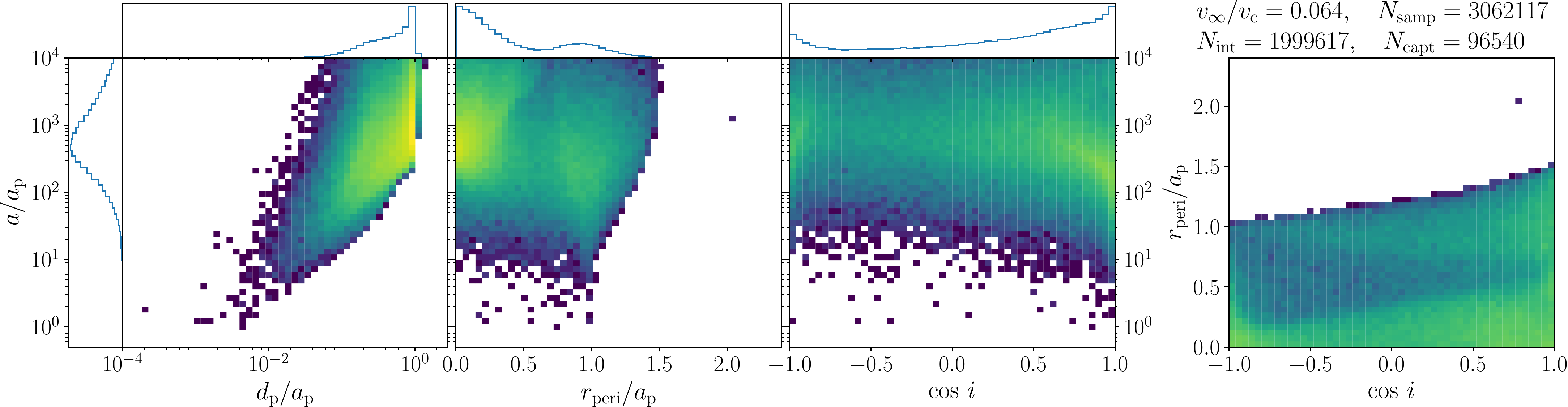}
	\vspace{1mm}
	\includegraphics[width=178.5mm]{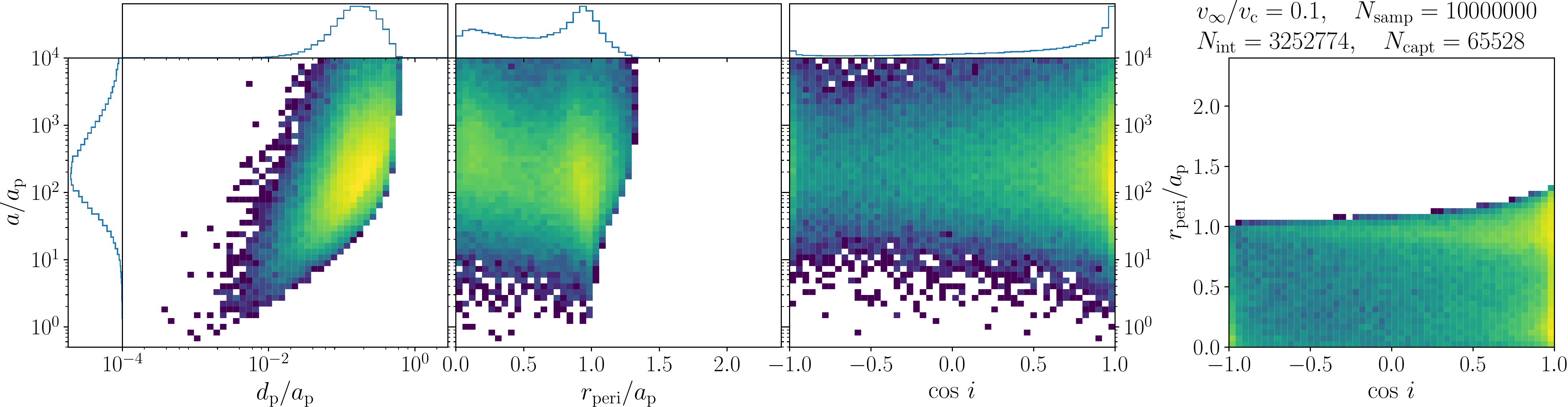}
	\caption{\label{fig:captured}
	Distribution over closest approach $\sub{d}p$ to the planet as well as semi-major axis $a$, perihelion distance $\sub{r}{peri}$ and (cosine of the) inclination $i$ for the simulated orbits captured at $a<10^4\sub{a}p$ from incoming speed $v_\infty$ as indicated (another value for each row) by a planet with mass $q=10^{-3}$. The 1D histograms (to the left an top) have a logarithmic scaling as is the colour scale used in the 2D histograms.
	Also given are the number of incoming orbits sampled, integrated, and captured. The figure continues on the next page.
	}
\end{figure*}
%%%%%%%%%%%%%%%%%
\setcounter{figure}{6}
\begin{figure*}
	\includegraphics[width=178.5mm]{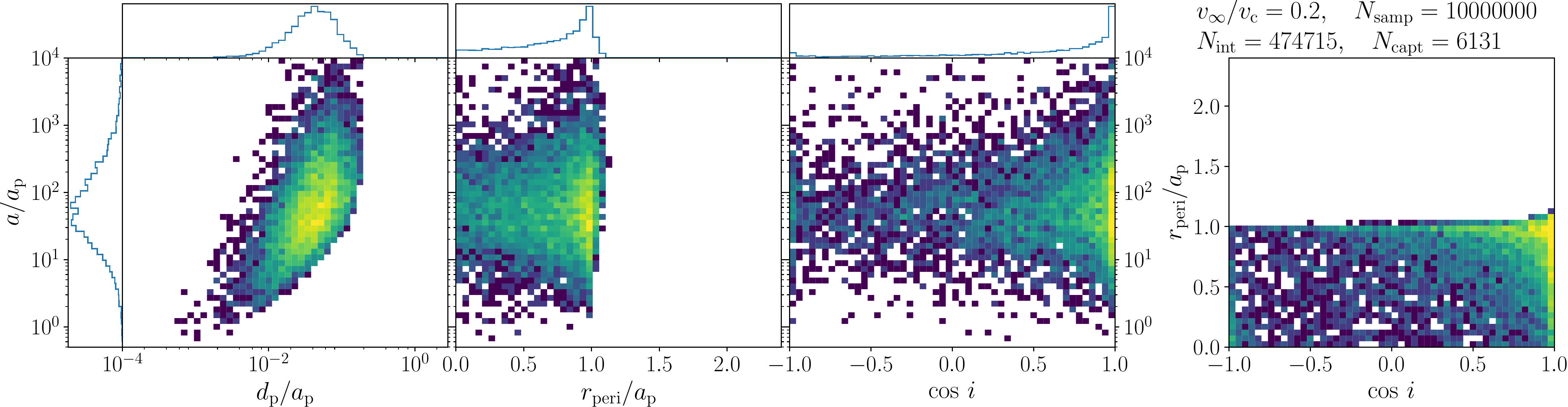}
	\vspace{1mm}
	\includegraphics[width=178.5mm]{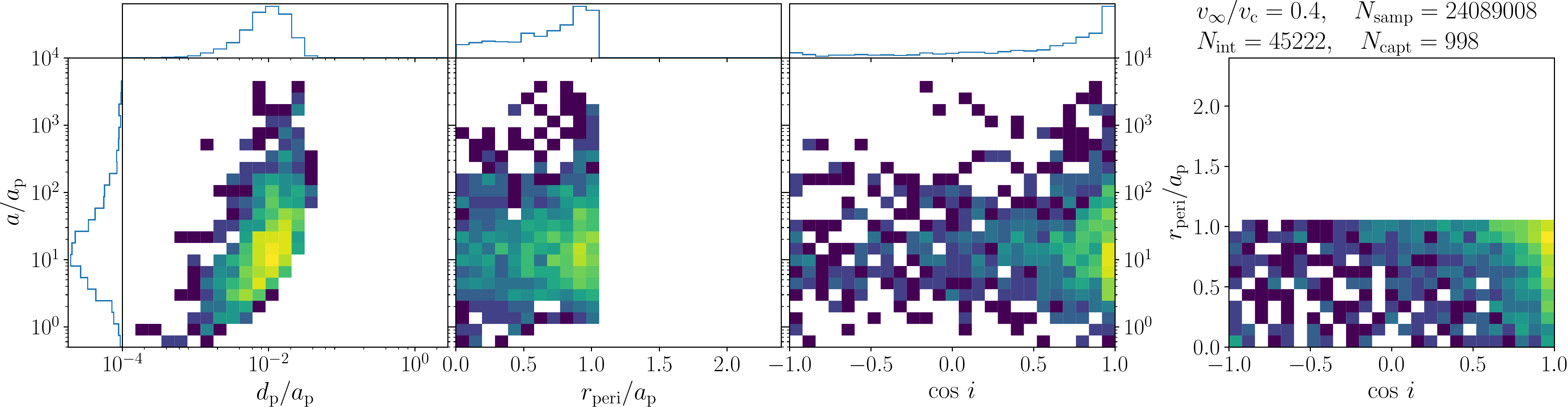}
	\vspace{1mm}
	\includegraphics[width=178.5mm]{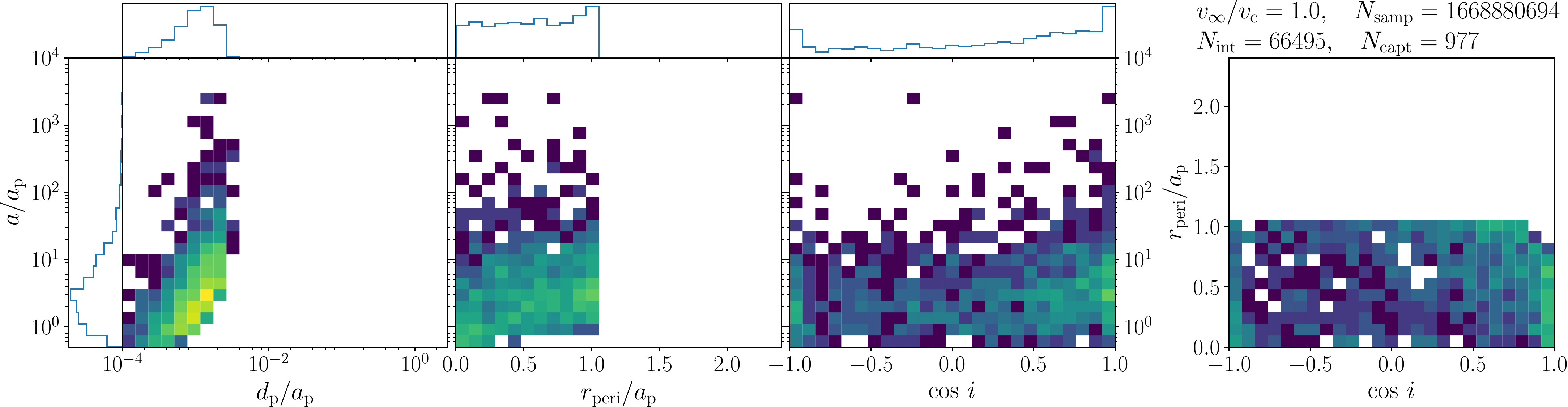}
	\caption{\label{fig:captured:ctd}
	    \textbf{continued}
	}
\end{figure*}
%%%%%%%%%%%%%%%%%

%%%%%%%%%%%%%%%%%%%%%%%%%%%%%%%%%%%%%%%%%%%%%%%%%%%%%%%
\subsubsection{Capture cross-section}
For each $v_\infty$ considered, we estimate the cumulative cross-sections $\sigma(v_\infty|\sub{v}a)$ for capture onto orbits with $\sub{E}{final}<-\frac12\sub{v}a^2$, i.e.\ semi-major axis $a<[\sub{v}c/\sub{v}a]^2\sub{a}p$, if at least 100 captured orbits satisfy this condition.

For the situation of a circular planet orbit ($\sub{e}p=0$) and various values of the mass ratio $q$, \autoref{fig:Sigma:X} plots the resulting cross-sections, multiplied by $(v_\infty/\sub{v}c)^2$, against the dimensionless measure $X$ (see equation~\ref{eq:X}) for the minimal energy change required to capture an ISO onto an orbit with semi-major axis $\le a$. Our theoretical considerations in Section~\ref{sec:analytic} suggest that the cumulative capture cross-section follows the general form
\begin{align}
    \label{eq:sigma:theory}
	\sigma(v_\infty|\sub{v}a) = \pi\sub{a}p^2\,\frac{\sub{v}c^2}{v_\infty^2}\,f(X)\,
	Y\left(1,\frac{v_\infty^2}{\sub{v}c^2},\frac{\sub{v}a^2}{\sub{v}c^2}\right).
\end{align}
Here, $Y$ is the transfer function of equation~\eqref{eq:Y}, which obtains $Y=1$ for $\sub{v}a,\,v_\infty\ll\sub{v}c$, i.e.\ for $X\lesssim q^{-1}$, and drops to zero at the maxima~\eqref{eq:ViVaMax}. At large $X$, the impulse approximation predicts $f=\tfrac83X^{-2}$ (equation~\ref{eq:sigma:capture}), while at small $X$, perturbation theory suggests the functional form $f\propto(s_{\max}/\sub{a}p)^2\sim|\ln X|^p$ (equation~\ref{eq:sigma:capture:wide}, see also \autoref{fig:Smax:X}). A simple function that combines these asymptotic limits is (thin curve in \autoref{fig:Sigma:X})
\begin{align}
    \label{eq:f(X):fit}
    \sub{f}{fit}(X) = \frac{8}{3X_0^2}\left[\sinh^{-1}(X_0/X)^{2/p}\right]^{p}
\end{align}
where a fit to the data at $X<q^{-1}$ obtains $X_0\approx2.95$ and $p\approx0.82$. 

The data also clearly show the drop of $\sigma$ at $X\gtrsim q^{-1}$ below this model due to the geometrical transfer function $Y$ and collisions with the planet (small symbols). The thin dotted lines in \autoref{fig:Sigma:X} give the prediction for $q=10^{-2}$ from equation~\eqref{eq:sigma:theory} for $f=\frac83X^{-2}$. They fit the data quite well, despite ignoring the finite size of the planet, i.e.\ collisions with the planet.

At $X\lesssim1$, the numerical results follow the expectation~\eqref{eq:sigma:capture:wide} from perturbation theory if we set the geometrical factor $\langle\xi\rangle\approx1/12$. The transition between these two regimes occurs at $X=X_0\approx3$ and is remarkably sharp. In particular, the relation $f(X)\approx\frac83X^{-2}$ from the impulse approximation holds down to much smaller values for $X$ than its formal range of validity $X\gtrsim 2q^{-1/3}$, as discussed in \Sect{sec:analytic:strong:limit}. This may just be a coincidence in the sense that the capture process at $1<X<2q^{-1/3}$, which is not accessible by analytic approximation, follows the same scaling. Indeed, the  small but systematic departures from $f=\frac83X^{-2}$ at $2<X<20$ (clearly visible in the bottom panel of \autoref{fig:Sigma:X}) indicate that some other process (than impulsive deflection) is at play here.

The functional form~\eqref{eq:sigma:theory} has a simple interpretation: $\pi\sub{a}p^2$ is the cross-section of the planetary orbit, which is enhanced by gravitational focusing accounted for by the factor $\sub{v}c^2/v_\infty^2$. The factor $f(X)$ is proportional to the probability that an incoming object coming close to the planet orbit suffers an energy change $\Delta E=-\frac12qX\sub{v}c^2$, while the factor $Y$ accounts for geometrical limits.

Finally, \autoref{fig:Sigma:X:ecc} shows the numerically estimated cross-section for different planetary eccentricities up to $\sub{e}p=0.4$ but fixed $q=10^{-3}$. The dependence of $\sigma$ on eccentricity is remarkably weak compared to its variation with $v_\infty$ or $\sub{v}a$ (i.e.\ $X$). In the regime of wide interactions $(X<1)$, the capture by planets on eccentric orbits is somewhat enhanced compared to circular orbits, while in the intermediate regime $1<X<10$ it is reduced. In the strong deflection limit (impulse approximation, $X>10$, the eccentricity dependence is at most very weak and below the uncertainties of the simulation data.

%%%%%%%%%%%%%%%%%%%%%%%%%%%%%%%%%%%%%%%%%%%%%%%%%%%%%%%
\subsubsection{Captured orbits}
For a planet with mass ratio $q=10^{-3}$ on a circular orbit, we plot in \autoref{fig:captured} the distributions over some characteristics of the orbits onto which the ISOs are captured for six of the 27 values for $v_\infty$ that we sampled (each row in the figure corresponds to another $v_\infty$). In particular, we show the 2D distributions of semi-major axis $a$ with the distance $\sub{d}p$ of closest approach to the planet, periapse radius and inclination, but also the 2D distribution between the latter two.

The first two values ($v_\infty/\sub{v}c=0.01$ and 0.03) are in the regime where capture is dominated by wide interactions. The orbital distributions in this case appear rather smooth and featureless. The maximal periapse radius $\sub{r}{peri}$ is much larger than $\sub{a}p$ and near-uniformly distributed below that. In the limit $e\sim1$ applicable here, this translates to a distribution uniform in $e^2$, i.e.\ ergodic. The distribution in inclination is also remarkably smooth and close to uniform, somewhat unexpected from our analytical arguments in \Sect{sec:analytic:capture:weak}, which suggested a prevalence of pro-grade orbits.

The last two rows (for $v_\infty/\sub{v}c=0.2$, 0.4 and 1) are firmly in the regime where capture requires close interactions with the planet. In this case, there is a preference for pro-grade orbits with $\sub{r}{peri}\sim\sub{a}p$, except for the highest values ($v_\infty\gtrsim\sub{v}c$), when the distribution returns to near uniform. The reason for this change is that the cross-section for capture depends correspondingly on the inclination of the incoming orbit (we have not considered such dependence in \Sect{sec:analytic:capture:strong}), but at very high $|\Delta E|$ a very close interaction is required in any case independent of inclination.

Most interesting is the intermediate regime of $q\lesssim (v_\infty/\sub{v}c)^2/\lesssim 10q$, where no analytic treatment was possible, and which is represented in \autoref{fig:captured} by the values $v_\infty/\sub{v}c=0.064$ and 0.1. In this case, the distribution in $\sub{r}{peri}$ is bi-modal, while pro-grade orbits become ever more likely with increasing $v_\infty$.

%%%%%%%%%%%%%%%%%%%%%%%%%%%%%%%%%%%%%%%%%%%%%%%%%%%%%%%
\section{Collision, Tidal Disruption, and Ejection}
\label{sec:coll+eject}
%%%%%%%%%%%%%%%%%%%%%%%%
\begin{table}
	\caption{
		\label{tab:collide}
		The rate at which major Solar-system bodies collide with ISOs is $\sub{\Gamma}{coll}=\sub{n}{iso}\sub{Q}{coll}$ with the \emph{volume capture rate} $\sub{Q}{coll}$ defined in equation~\eqref{eq:Q:coll}. Similarly, the rate at which cometary ISOs (with density 0.6\,g\,cm$^{-3}$) are tidally disrupted by close encounters is $\sub{\Gamma}{tid}=\sub{n}{iso}\sub{Q}{tid}$ with $\sub{Q}{tid}$ computed analogously with $\sub{\sigma}{coll}$ replaced by $\sub{\sigma}{tid}$ of equation~\eqref{eq:sigma:tid}.
	}
	\centering
	\begin{tabular}{lll}
	planet/moon &
	$\sub{Q}{coll}$ [au$^3$\,yr$^{-1}$] &
	$\sub{Q}{tid}$ [au$^3$\,yr$^{-1}$] \\
	\hline
	Sun & 0.1737 & 3.088 \\
	\hline
	Mercury & $3.82 \times 10^{-8}$ & $2.62 \times 10^{-5}$ \\
	Venus 	& $1.53 \times 10^{-7}$ & $1.00 \times 10^{-4}$ \\
	Earth 	& $1.40 \times 10^{-7}$ & $9.42 \times 10^{-5}$ \\
	Mars 	& $3.09 \times 10^{-8}$ & $1.70 \times 10^{-5}$ \\
	Jupiter & $2.52 \times 10^{-5}$ & $2.57 \times 10^{-3}$ \\
	Saturn 	& $9.55 \times 10^{-6}$ & $9.85 \times 10^{-4}$ \\
	Uranus 	& $1.24 \times 10^{-6}$ & $2.55 \times 10^{-4}$ \\
	Neptune & $1.20 \times 10^{-6}$ & $2.78 \times 10^{-4}$ \\
	\hline
	Moon    & $1.94 \times 10^{-8}$ & $5.99 \times 10^{-6}$ \\
	Io      & $1.57 \times 10^{-8}$ & $4.85 \times 10^{-6}$ \\
	Europa  & $9.13 \times 10^{-9}$ & $2.81 \times 10^{-6}$ \\
	Ganymede& $2.03 \times 10^{-8}$ & $6.27 \times 10^{-6}$ \\
	Callisto& $1.11 \times 10^{-8}$ & $3.44 \times 10^{-6}$ \\
	Titan   & $1.11 \times 10^{-8}$ & $3.41 \times 10^{-6}$ \\
	Triton  & $2.82 \times 10^{-9}$ & $8.70 \times 10^{-7}$ \\
	\hline
	planets \& moons
	& $3.77 \times 10^{-5}$ & $4.36 \times 10^{-3}$ \\
	\hline
	\end{tabular}
\end{table}
%%%%%%%%%%%%%%%%%%%%%%%%

%%%%%%%%%%%%%%%%%%%%%%%%%%%%%%%%%%%%%%%%%%%%%%%%%%%%%
\subsection{Collisions and Tidal disruptions of ISOs}
One byproduct of the analytical treatment of close encounters with the planets was the derivation of the cross-section for collisions of ISOs with planets or moons in equations~\eqref{eq:sigma:collision} and~\eqref{eq:sigma:collision:moon}, respectively. Integrating over the ISO velocity distribution, obtains the collision rate as $\sub{\Gamma}{coll}=\sub{n}{iso}\sub{Q}{coll}$ with the \emph{volume collision rate} 
\begin{align}
	\label{eq:Q:coll}
	\sub{Q}{coll} = \int_0^\infty\diff v_\infty\; v_\infty\,p(v_\infty)\,\sub{\sigma}{coll}(v_\infty),
\end{align}
where $p(v_\infty)$ is the distribution of ISO asymptotic speeds. Assuming that $p(v_\infty)$ is the same as the distribution observed for stars in the Solar neighbourhood from Gaia (see \citetalias{paper2} for details) obtains the numbers reported in Table~\ref{tab:collide} for the Solar system planets. With $\sub{n}{iso}\sim0.1\,$au$^{-3}$ we find that Earth is hit by an ISO only once every 70\,Myr.

ISOs that just avoid collision may get tidally disrupted if their closest approach distance $\sub{d}p$ to the (centre of the) planet is smaller than their Roche limit
\begin{equation}
    \label{eq:d:tid}
    \sub{d}{tid} \approx \sub{R}p \left(2\frac{\sub{\rho}{p}}{\sub{\rho}{iso}}\right)^{1/3}.
\end{equation}
The conditions for tidal disruption are therefore that $\sub{\rho}{iso}<2\sub{\rho}p$ and $\sub{d}p < \sub{d}{tid}$. From the distribution of simulated captures over the closest distance $\sub{d}p$ to the planet in \autoref{fig:captured}, we see that only a few ISOs with incoming speeds $v_\infty\sim\sub{v}c$ reach $\sub{d}p\le1.4\times10^{-4}\sub{a}p$, which is the Roche limit for $\sub{\rho}{iso}=0.5\sub{\rho}{p}$, corresponding to comets passing Jupiter. Hence, tidal disruption of captured ISOs is extremely rare.

The cross-section $\sub{\sigma}{tid}$ for ISO's to be tidally disrupted by a close encounter is hence that of having a closest approach to the planet satisfying $\sub{R}p < \sub{d}p < \sub{d}{tid}$. Using our results for the collision cross-section, this can be obtained as
\begin{align}
    \label{eq:sigma:tid}
    \sub{\sigma}{tid}(v_\infty) = 
    \max\left\{0,
    \sub{\sigma}{coll}(v_\infty|\sub{R}p\to\sub{d}{tid}) -
    \sub{\sigma}{coll}(v_\infty)
    \right\},
\end{align}
where the replacement $\sub{R}p\to\sub{d}{tid}$ includes the relation for the escape speed. Table~\ref{tab:collide} gives the corresponding volume rates $\sub{Q}{tid}$ in column three (again assuming that ISOs have the same speed distribution as nearby stars).

The distinction between tidal disruption events by planets or moons as opposed to the Sun is relevant inasmuch as slightly less than half of these occur during close encounters that capture the ISO or rather its fragments into the Solar system (while no such capture occurs for tidal disruption by the Sun). For cometary ISOs, $\sub{n}{iso}\sim0.009\,$au$^{-3}$ \citep{SirajLoeb2021} and we expect $\sim20$ of them to be tidally disrupted and their fragments captured per Myr.

%%%%%%%%%%%%%%%%%
\begin{figure}
	\includegraphics[width=\columnwidth]{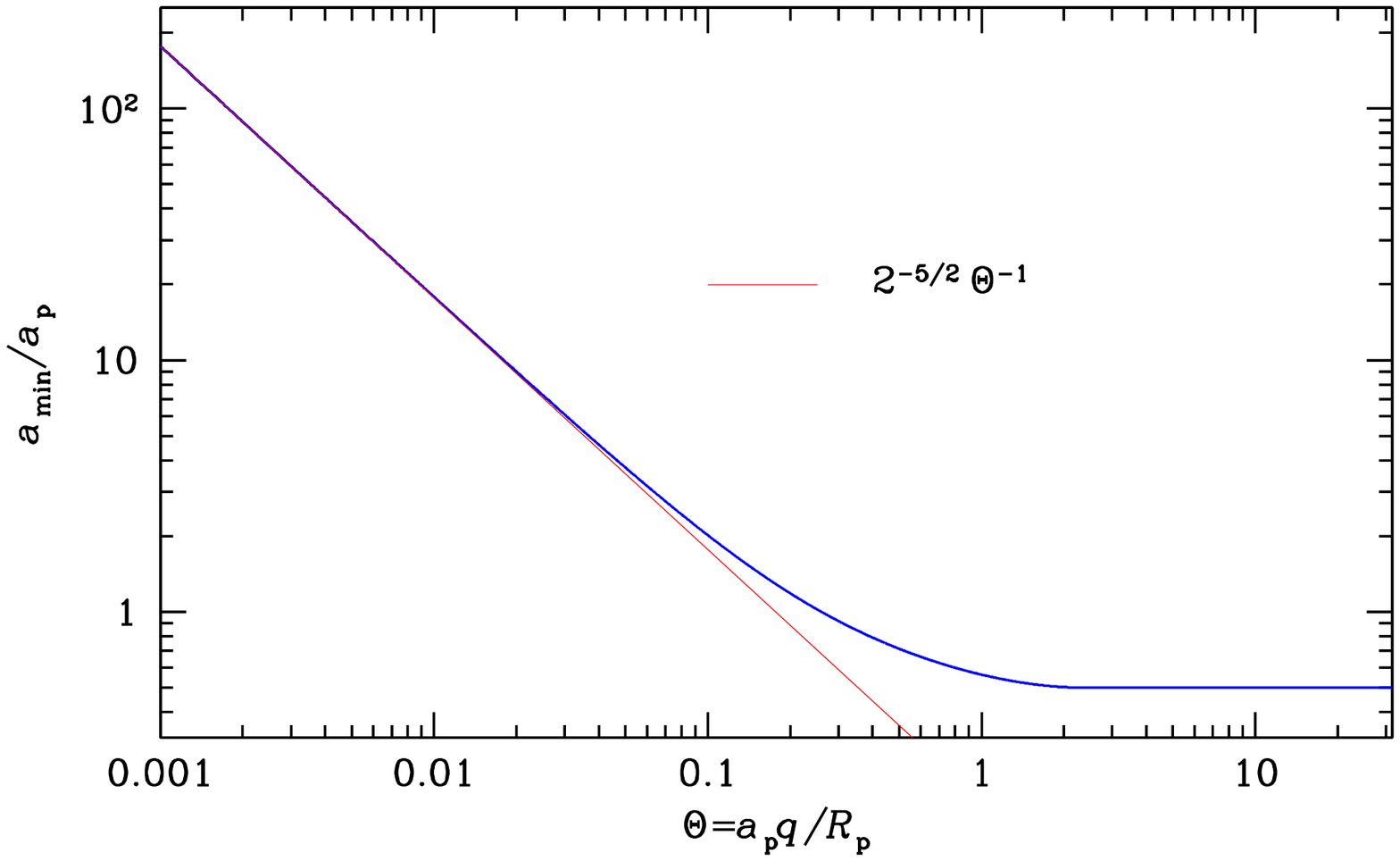}
	\caption{\label{fig:Theta:0}
	A planet with radius $\sub{R}p$ on a circular orbit with semi-major axis $\sub{a}p$ and Safronov number $\Theta$ can eject via a single slingshot any object with semi-major axis $a\ge a_{\min}$ (blue). For compact planets ($\Theta\ge1+\sqrt{2}\approx2.414$) $a_{\min}=\sub{a}p/2$, while for large planets $a_{\min}=2^{-5/2}\sub{R}p/q$ (red).
	}
\end{figure}
%%%%%%%%%%%%%%%%%
%%%%%%%%%%%%%%%%%
\begin{figure}
	\includegraphics[width=\columnwidth]{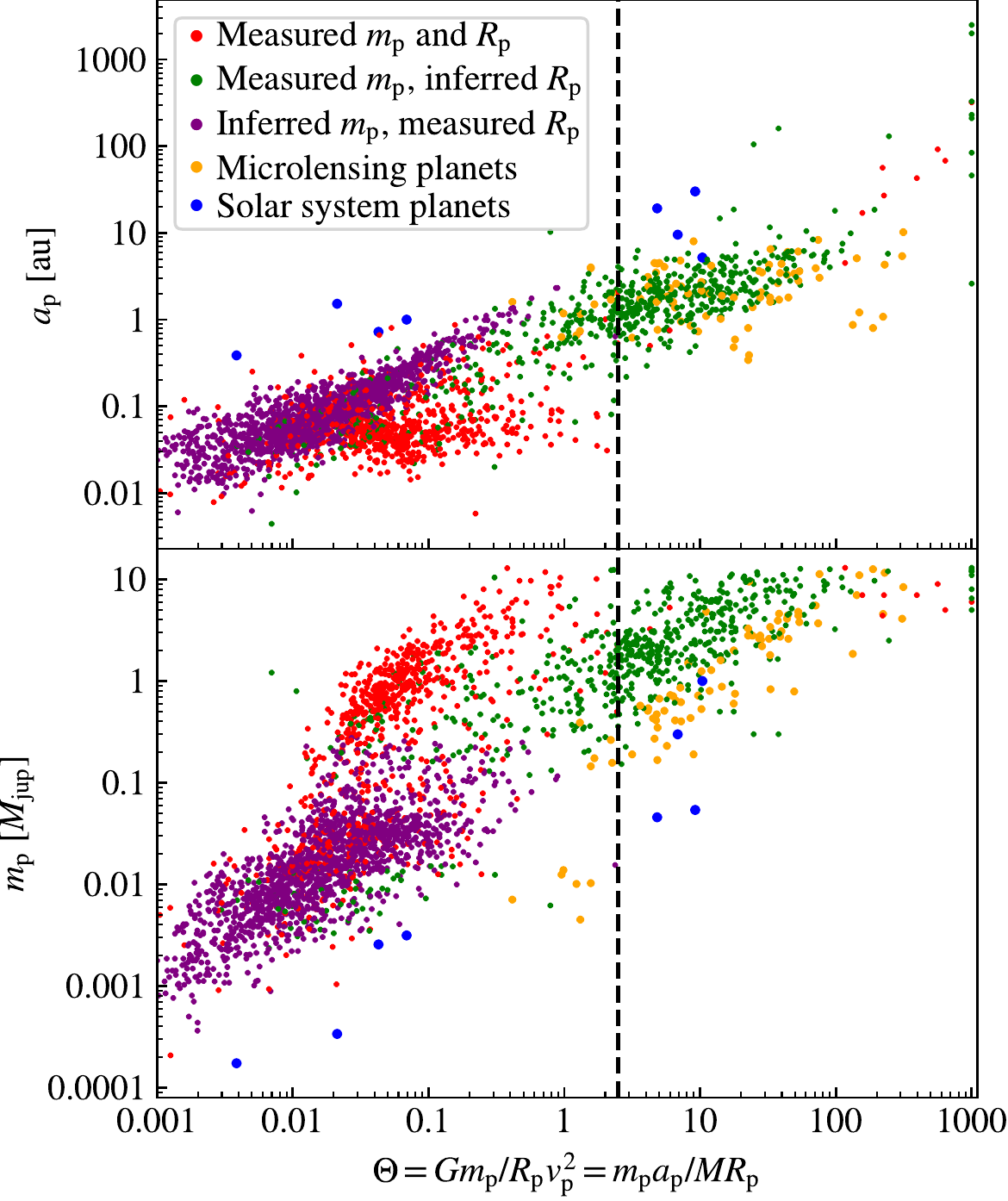}
	\vspace*{-5mm}
	\caption{\label{fig:Theta}
	Distribution of solar-system planets (blue) and exoplanets over semi-major axis $\sub{a}p$, mass $\sub{m}p$, and Safronov number $\Theta$ (clamped to $\Theta\le1000$), assuming circular orbits. Missing measurements for $\sub{R}p$ and $\sub{m}p$ were replaced by values inferred from $\sub{R}p(\sub{m}p)$ relations by \citet{Otegi2020} for $\sub{m}p < 124 M_{\earth}$ and \citet{Bashi2017} for more massive planets (excluding $0.18< \sub{R}p/\sub{R}{jup} <0.27$ where these relations are not unique). Only compact planets with $\Theta>2.414$ (dashed) can eject any orbit-crossing bound object via a single slingshot, while larger objects can only eject objects with ever greater semi-major axes, see also \autoref{fig:Theta:0}.}
\end{figure}
%%%%%%%%%%%%%%%%%

%%%%%%%%%%%%%%%%%%%%%%%%%%%%%%%%%%%%%%%%%%%%%%%%%%%%%
\subsection{Sling-shot ejections of asteroids}
Another byproduct of the treatment of close interactions was the maximum capturable speed $v_\infty$ and binding energy $GM/a=\tfrac12\sub{v}a^2$ for the captured ISO as function of the planet radius $\sub{R}{p}$ through the Safronov number $\Theta=q\sub{a}p/\sub{R}p$. These limits also apply to the reverse process: ejection of a bound object by a single slingshot. In \autoref{fig:Theta:0} we plot the resulting minimum semi-major axis (corresponding to the maximum binding energy) of a bound object that a planet with given Safronov number $\Theta$ can just eject. For $\Theta<1+\sqrt{2}\approx2.4$ the ability of a planet to eject orbit crossing objects is diminished. However, weakly bound orbits with semi-major axes twice that of the planet or larger can still be ejected via a single slingshot by a large planet with Safronov number as small as $\Theta=0.1$, hundred times smaller than for Jupiter.

\autoref{fig:Theta} shows the Safronov numbers $\Theta$ for exoplanets with measured/inferred radius and mass. Obviously most of these have $\Theta\lesssim0.1$, implying that they are incapable to eject most orbit crossing planetesimals in a single slingshot. However, this is of course largely owed to the selection inherent to the detection methods, which favour large planets with small semi-major axes.

Of course, these planets may still eject bound objects (rather than collide with them) via multiple wide encounters, slowly increasing the semi-major axis of the objects until ejection. This process may be slow, but planets on orbits $\sim100$ times smaller than Jupiter's have $\sim1000$ times more orbital periods to act.

%%%%%%%%%%%%%%%%%%%%%%%%%%%%%%%%%%%%%%%%%%%%%%%%%%%%%%%
\section{Conclusion}
\label{sec:conclude}
%%%%%%%%%%%%%%%%%%%%%%%%%%%%%%%%%%%%%%%%%%%%%%%%%%%%%%%
This study revisited the capture of initially unbound interstellar objects (ISOs) into bound states of a planet-star binary with semi-major axis $\sub{a}p$ in general. We first considered in \Sect{sec:analytic} analytical arguments based on the impulse approximation for close encounters with the planet (extending previous studies) and on perturbation theory for wide encounters.

In \Sect{sec:sigma:sim} we performed numerical simulations, whereby substantially extending previous studies also to planets with eccentricity $\sub{e}p>0$. These simulations confirmed the analytical results for the asymptotic cases when the energy change $|\Delta E|$ required for capture is much larger (close encounters) or much smaller (wide encounters) than $q\sub{v}c^2=qGM/\sub{a}p$. The transition between these two well understood asymptotic regimes at $|\Delta E|\sim q\sub{v}c^2$ is remarkably sharp.

The general result of these investigations is that the capture cross-section $\sigma$ is a steeply declining function of both the incoming speed $v_\infty$ of the ISO and the energy change $|\Delta E|$ (equation~\ref{eq:sigma:theory}). $\sigma$ depends on $|\Delta E|$ and the planet-to-star mass ratio $q$ only via their ratio $X\equiv|2\Delta E|/q\sub{v}c^2$, except that it drops to zero at $X\gtrsim q^{-1}$, corresponding to the closest encounters possible. This truncation is due to orbital geometry: the maximum captureable incoming speed is $v_\infty\approx3\sub{v}c$, even for point-like planets. In reality, the planet's finite size limits the already tiny probability to capture objects with $v_\infty\gtrsim\sub{v}c$ further, since the corresponding slingshot approach the planet closer than its radius $\sub{R}p$. The effect of the planet size is most naturally expressed via its Safronov number $\Theta\equiv q\sub{a}p/\sub{R}p$, which for Jupiter and Saturn is sufficiently large for their capture cross-sections to be largely unaffected by collisions.

The capture cross-section depends only weakly on the planet's orbital eccentricity $\sub{e}p$, the most notable effect being an increase of $\sigma$ by a factor $\sim1+\sub{e}p^2$ near $X=0.05$, when captures are dominated by wide encounters. The resulting capture rate is treated in an accompanying paper \citet{paper2} with particular emphasis on the Solar system, including the resulting steady-state population of captive ISOs.

%%%%%%%%%%%%%%%%%%%%%%%%%%%%%%%%%%%%%%%%%%%%%%%%%%%%%%%
\section*{Acknowledgements}
%%%%%%%%%%%%%%%%%%%%%%%%%%%%%%%%%%%%%%%%%%%%%%%%%%%%%%%
We thank Douglas Heggie for discussions that lead to improving Sections~\ref{sec:analytic} \&~\ref{sec:sigma:sim}, Ralph Sch\"onrich for many stimulating conversations, Ravit Helled for helpful exchanges, and the reviewer, Simon Portegies Zwart, for useful comments. TH was supported by a University of Z\"urich Forschungskredit (grant K-76102-02-01). Part of this work has been carried out in the frame of the National Centre for Competence in Research ‘PlanetS’ supported by the Swiss National Science Foundation (SNSF) (grant F-76102-04-01). The simulations of \Sect{sec:sigma:sim} were performed using the DiRAC Data Intensive service at Leicester, operated by the University of Leicester IT Services, which forms part of the STFC DiRAC HPC Facility (\url{www.dirac.ac.uk}). The equipment was funded by BEIS capital funding via STFC capital grants ST/K000373/1 and ST/R002363/1 and STFC DiRAC Operations grant ST/R001014/1. DiRAC is part of the National e-Infrastructure.

%%%%%%%%%%%%%%%%%%%%%%%%%%%%%%%%%%%%%%%%%%%%%%%%%%%%%%%
\section*{Data Availability}
%%%%%%%%%%%%%%%%%%%%%%%%%%%%%%%%%%%%%%%%%%%%%%%%%%%%%%%
The results of the numerical simulation of \Sect{sec:sigma:sim} are available on reasonable request to the corresponding author.

%%%%%%%%%%%%%%%%%%%%%%%%%%%%%%%%%%%%%%%%%%%%%%%%%%%%%%%
\bibliographystyle{mnras} \bibliography{capture}
%%%%%%%%%%%%%%%%%%%%%%%%%%%%%%%%%%%%%%%%%%%%%%%%%%%%%%%
\appendix
%%%%%%%%%%%%%%%%%%%%%%%%%%%%%%%%%%%%%%%%%%%%%%%%%%%%%%%
\section{Slingshot ejection}
\label{app:eject}
Here, we derive the ejection rate of captured ISOs via slingshots by the planet, which is needed in \citetalias{paper2}, but technically fits better here. The process is the reverse of the capture mechanism treated in \Sect{sec:analytic:capture:strong} and the calculation of the flux proceeds analogously. However, the initial conditions are different: in \Sect{sec:analytic:capture:strong} we considered the scattering of an incoming population of ISOs with singular (specific) energy $E=v_\infty^2/2$ onto bound orbits with a range of energies $E\le-\sub{v}a^2/2$, while here we instead consider the scattering of a bound population with singular $E=-\sub{v}a^2/2$ onto unbound orbits in the range $E\ge v_\infty^2/2$. 

%%%%%%%%%%%%%%%%%%%%%%%%%%%%%%%%%%%%%%%%%%%%%%%%%%%%%%%
\subsection{The flux of ejections}
When the object enters and exits the planet Roche sphere, it has barycentric speeds
\begin{align}
    \label{eq:ej:v12}
    v_1^2 = 2\sub{v}r^2 - \sub{v}a^2,  \qquad
    v_2^2 = 2\sub{v}r^2 + v_\infty^2,
\end{align}
and planetocentric speed $w$ given by equation~\eqref{eq:w:alpha}. Relations~\eqref{eqs:v12:theta} also hold here, and the ejection condition is obtained by inserting $v_2^2-v_1^2=v_\infty^2+\sub{v}a^2$:
\begin{align}
	\label{eq:ejct:condition:0}
	v_\infty^2 + \sub{v}a^2 = -4G\sub{m}p\sub{v}pw
	\frac{G\sub{m}p\cos\theta+\sub{b}pw^2\sin\theta\cos\sub{\psi}p}{G^2\sub{m}p^2 + \sub{b}p^2 w^4},
\end{align}
which differs from the capture condition~\eqref{eq:capt:condition:0} only by the sign of the energy change. Combining equations~\eqref{eq:v:1:theta} and~\eqref{eq:ej:v12} we find $-2\sub{v}pw\cos\theta=D$ (as defined in equation~\ref{eq:D}) and can eliminate $\theta$ from equation~\eqref{eq:ejct:condition:0} to obtain the ejection condition for $w$, $\sub{b}p$, and $\sub{\psi}p$ at given $v_\infty$ and $\sub{v}a$:
\begin{align}
	\label{eq:ejct:condition:1}
	v_\infty^2 + \sub{v}a^2 = \frac{2G^2\sub{m}p^2}{G^2\sub{m}p^2+\sub{b}p^2w^4}
	\left[D-\cos\sub{\psi}p\frac{\sub{b}pw^2}{G\sub{m}p}
	\sqrt{4\sub{v}p^2w^2-D^2}\right].
\end{align}
The solution for $\sub{b}p$ and $\sub{\psi}p$ of this condition at given $v_\infty$ and $\sub{v}a$ lies on a circle~\eqref{eq:capt:condition:2} in the impact plane with offset and radius
\begin{subequations}
\label{eqs:ejct:b0,rho}
\begin{align}
	\label{eq:ejct:b0}
	b_0 &=-\frac{G\sub{m}p}{(v_\infty^2 + \sub{v}a^2)w^2}
	\sqrt{4w^2\sub{v}p^2-D^2},
	\\
	\label{eq:ejct:rho}
	\rho &= \phantom{-} \frac{G\sub{m}p}{(v_\infty^2 + \sub{v}a^2)w^2}
	\sqrt{4w^2\sub{v}p^2-C^2}.
\end{align}
\end{subequations}
Compared to the corresponding relations~\eqref{eqs:b0,rho} for capture, $b_0$ and $\rho$ have swapped, which is owed to the different initial conditions relating to $v_\infty$ for capture and to $\sub{v}a$ for ejection. In addition, the sign of $b_0$ is changed, because an energy increase (required for ejection) is favoured by orbits passing behind the planet, while for capture passing in front of the planet is better.

\begin{figure}
    \begin{center}
	\includegraphics[width=85mm]{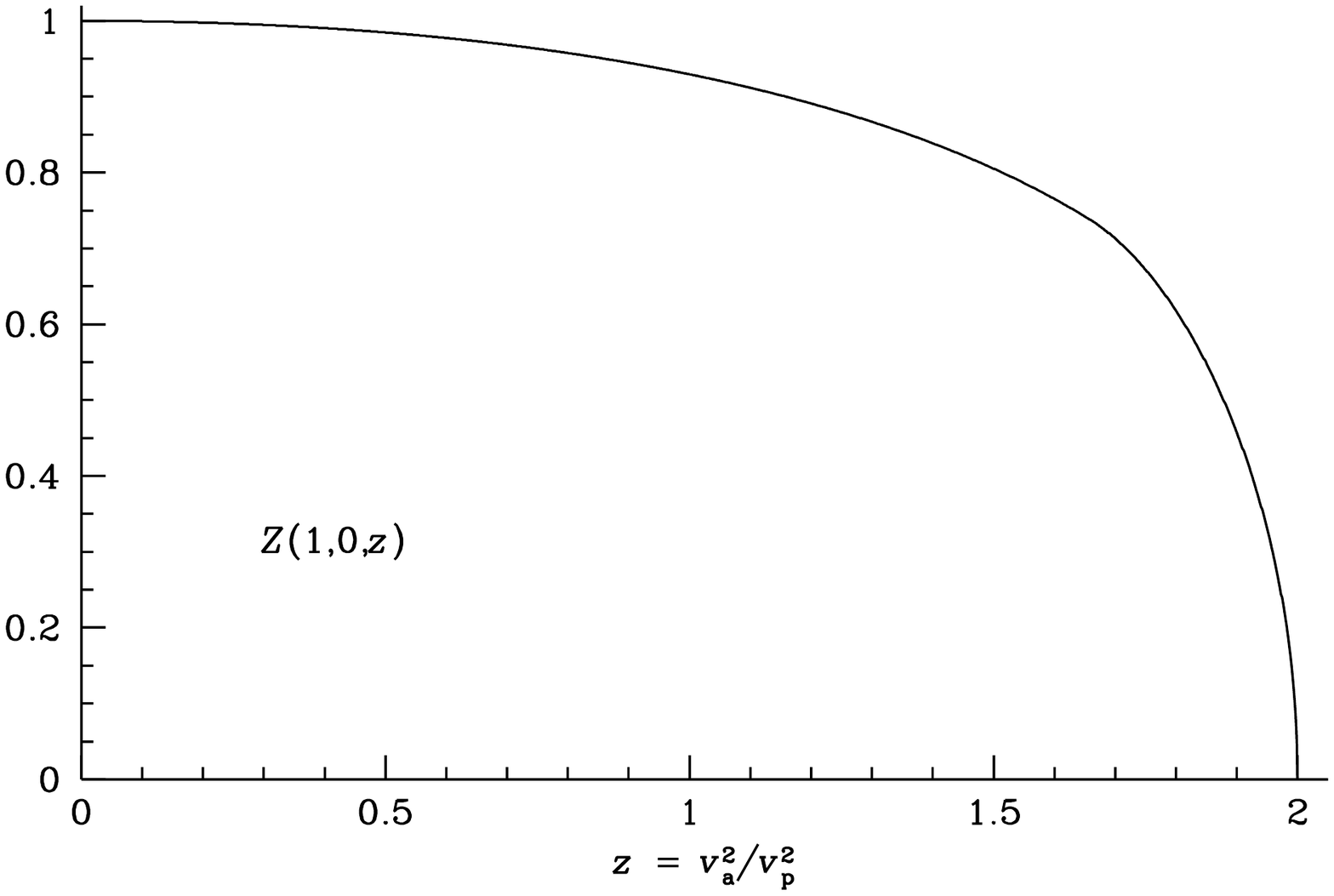}
    \end{center}
	\caption{\label{fig:Z0}
	The transfer function $Z(1,0,z)$ defined in equation~\eqref{eq:Z}, which governs all ejections from a planet on a circular orbit.
	}
\end{figure}

We proceed analogously to the treatment of capture and average the local flux $F=n_1w\pi\rho^2$ of objects scattered by the planet onto unbound orbits over all orientations $\alpha$ of the incoming orbit using equation~\eqref{eq:dcosalpha} with the limits~\eqref{eq:w:limits}. This gives
\begin{align}
	\label{eq:ejct:F:alpha}
	\langle F\rangle_\alpha
	\doteq \frac{8\pi}{3}
		\frac{n_1\,G^2\sub{m}p^2\sub{v}p^2}{(v_\infty^2 + \sub{v}a^2)^2v_1}\,
		Z\left(\frac{1}{1+\sub{e}p\cos\sub{\eta}p},\frac{v_\infty^2}{\sub{v}p^2},\frac{\sub{v}a^2}{\sub{v}p^2}\right),
\end{align}
with the transfer function
\begin{align}
	\label{eq:Z}
	Z(u,x,z) &\equiv 
		\left[\frac{3}{16y}\left(2u-1+x\right)^2
	  + \frac{3y}{8}\left(2u+1+x\right)
	  - \frac{y^3}{16}\right]_{y_-}^{y_+},
\end{align}
where $y_-$ and $y_+$ are the same as for the capture transfer function~\eqref{eq:Y}. Equation~\eqref{eq:ejct:F:alpha} for the averaged flux $\langle F\rangle_\alpha$ differs from that for capture only by the transfer function. The transfer function $Z$ for ejection has of course the same domain as the transfer function $Y$ for capture but is everywhere larger (except at $x,z=0,0$ and on the edge of the domain), see also \autoref{fig:Z0}.

%%%%%%%%%%%%%%%%%
%%%%%%%%%%%%%%%%%
Since here we are not interested in the asymptotic speed $v_\infty$ of the ejected object, we set $v_\infty=0$ in equation~\eqref{eq:ejct:F:alpha} and also assume a circular planet orbit when the local cross-section for ejection becomes
\begin{align}
	\label{eq:ejct:F:alpha:0}
	\sub{\sigma}{ej}(a)
	\doteq \frac{8\pi}{3}
		\frac{a^2q^2}{2-\sub{a}p/a}
		Z\left(1,0,\frac{\sub{a}p}{a}\right).
\end{align}
For most of the relevant range $Z(1,0,z)\approx1$, as shown in \autoref{fig:Z0}, and we will in the following use this approximation.

%%%%%%%%%%%%%%%%%%%%%%%%%%%%%%%%%%%%%%%%%%%%%%%%%%%%%%%
\subsection{The retention time of captured ISOs}
ISOs which have been captured by a slingshot with the planet will cross into radii $r<\sub{a}p$ once per orbit and have the chance to suffer a slingshot with the planet. The probability to be ejected is therefore $\sim\sub{\sigma}{ej}/\pi\sub{a}p^2$ per orbit. The typical time an ISO remains captive is then its orbital period divided by this probability
\begin{align}
    \label{eq:T:captive}
    \sub{T}{stay}(a) \approx \frac3{4q^2} \sqrt{\frac{\sub{a}p}{a}}
    \left[1-\frac{\sub{a}p}{2a}\right]\sub{T}p
\end{align}
where $\sub{T}p=2\pi/\sub{\Omega}p$ is the orbital period of the planet.

%%%%%%%%%%%%%%%%%%%%%%%%%%%%%%%%%%%%%%%%%%%%%%%%%%%%%%%
\section{Numerical Orbit integration}
\label{app:integrator}
For the numerical orbit integrations in section~\ref{sec:sigma:sim:orbit} we use the symplectic time-reversible integrator introduced by \citeauthor{HernandezBertschinger2015} (\citeyear{HernandezBertschinger2015}, see also \citealt{DehnenHernandez2017}), which splits the Hamiltonian into two-body Hamiltonians (to be integrated by a Kepler solver) and the surplus kinetic energies (resulting in backwards drifts).
In what follows, we use subscripts `s', `p', and `e' for star, planet, and exobody, respectively. The Hamiltonian of the three-body system can be split into that of the star-planet binary and the energy of the ISO: $H=\sub{H}{sp}+\sub{m}e\sub{H}{e}$ with \begin{align}
	\sub{H}{sp}  &= \frac{\sub{\vec{p}}s^2}{2\sub{m}s} 
				  + \frac{\sub{\vec{p}}p^2}{2\sub{m}p}
				  - \frac{G\sub{m}s\sub{m}p}{\sub{r}{sp}},
	\\
	\sub{H}{e}   &= \frac{\sub{\vec{v}}e^2}{2} - \frac{G\sub{m}s}{\sub{r}{es}}
				  - \frac{G\sub{m}p}{\sub{r}{ep}}.
\end{align}
Here, $\sub{\vec{v}}e=\sub{\vec{p}}e/\sub{m}e$ is the specific momentum (velocity) of the ISO, while $\vec{x}_{i\!\!j}\equiv\vec{x}_i-\vec{x}_{\!j}$ and $r_{i\!\!j}\equiv|\vec{x}_{i\!\!j}|$. In our simulations, we let $\sub{m}e\to0$, corresponding to the restricted three-body problem. In this limit, the contribution of $\sub{H}{e}$ to $H$ vanishes and our integrator, which constructs the particle trajectories as piecewise Kepler orbits, integrates the binary motion exactly, but not the ISO motion. Rather, the numerically integrated ISO orbit corresponds to the surrogate Hamiltonian \citep{DehnenHernandez2017}
\begin{align}
	\label{eq:H:ISO:surr}
	\sub{\tilde{H}}{e} &= \sub{H}{e} + 
	\tfrac{1}{24}\tau^2 \sub{H}{e,2} + O(\tau^4),
\end{align}
where $\tau$ denotes the time step and
\begin{align}
	\sub{H}{e,2} &\equiv G^2\sub{m}s\sub{m}p \left[
		\frac{\sub{\vec{x}}{ep}\cdot\sub{\vec{x}}{es}}{\sub{r}{ap}^3\sub{r}{es}^3} -
		\frac{\sub{\vec{x}}{ep}\cdot\sub{\vec{x}}{sp}}{\sub{r}{ep}^3\sub{r}{sp}^3} -
		\frac{\sub{\vec{x}}{es}\cdot\sub{\vec{x}}{sp}}{\sub{r}{as}^3\sub{r}{sp}^3}	
	\right].
\end{align}
Since the trajectory is modelled as alternating Kepler orbits around star and planet, close encounters with these are exactly integrated and $\sub{\tilde{H}}{e}$ contains only terms to which all three particles contribute, i.e.\ errors arise only from three-body encounters. The $O(\tau^2)$ error term $\sub{H}{e,2}$ can vary considerably along the orbit and we therefore adapt $\tau$ using the scheme of \cite{HairerSoderlind2005}, which amounts to reversibly integrating the integration frequency $1/\tau$ such that $\tau$ closely follows a time-step function $T$. This method is no longer exactly symplectic, but still time reversible and very efficient for our problem.

A suitable choice for $T$ would be $T\propto\sqrt{\eta/|\sub{H}{e,2}|}$ with some accuracy parameter $\eta$, such that the local error remains constant. In order to avoid $T\to\infty$, we replace $\sub{H}{e,2}$ with the positive definite
\begin{align}
	\label{eq:E:2}
	\sub{E}{e,2} \equiv 
		G^2\sub{m}s\sub{m}p \left[
		\frac{1}{\sub{r}{ep}^2\sub{r}{es}^2} +
		\frac{1}{\sub{r}{ep}^2\sub{r}{sp}^2} +
		\frac{1}{\sub{r}{es}^2\sub{r}{sp}^2}	
	\right] \ge |\sub{H}{e,2}|.
\end{align}
Unfortunately, the distances $\sub{r}{ep}$ and (in case of an elliptic binary) $\sub{r}{sp}$ change on the planet's orbital time scale, which may be shorter than $\tau$ when the ISO is still far from the binary.\footnote{This is also the reason why a fourth-order version of the integrator, which simply integrates the error term $\sub{H}{e,2}$ \citep{DehnenHernandez2017}, fails to obtain substantially better accuracy, as measured by the conservation of Jacobi's integral for circular planet orbits.} As a result $|\dot{T}|\gtrsim1$ and an accurate time integration of $1/\tau$ fails in this situation. To avoid this problem, we replace $\sub{r}{ep}$ and $\sub{r}{sp}$ in equation~\eqref{eq:E:2} by
\begin{align}
	\sub{\tilde{r}}{ep} = \sub{r}{ep}^{1-\alpha}\sub{r}{es}^{\alpha}
	\qquad\text{and}\qquad
	\sub{\tilde{r}}{sp} = \sub{r}{sp}^{1-\alpha}\sub{a}p^{\alpha}
\end{align}
respectively, with $\sub{a}p$ the binary's semi-major axis and $\alpha = \sub{r}{es}^2/(\sub{r}{es}^2+100\sub{a}p^2)$,
%\begin{align}
%	\alpha = \frac{\sub{r}{es}^2}{\sub{r}{es}^2+100\sub{a}p^2},
%\end{align}
such that $\tilde{r}_{i\mathrm{p}} \to r_{i\mathrm{p}}$ as the ISO approaches the binary.

Finally, we set $T=\sqrt{\eta E/\sub{E}{e,2}}$ with some reference energy $E$. A suitable value for $E$ is the Jacobi integral of a parabolic orbit with specific angular momentum comparable to that of the binary, i.e.\ $E=\Omega^2\sub{a}p^2=G(\sub{m}s+\sub{m}p)/\sub{a}p$, when
\begin{align}
	\label{eq:T}
	T^{-2} = \frac1\eta \frac{G\sub{m}s\sub{m}p}{\sub{m}s+\sub{m}p}
	\left[
		\frac{\sub{a}p}{\sub{\tilde{r}}{ep}^2\sub{r}{es}^2} +
		\frac{\sub{a}p}{\sub{\tilde{r}}{ep}^2\sub{\tilde{r}}{sp}^2} +
		\frac{\sub{a}p}{\sub{r}{es}^2\sub{\tilde{r}}{sp}^2}	
	\right].
\end{align}

%%%%%%%%%%%%%%%%%%%%%%%%%%%%%%%%%%%%%%%%%%%%%%%%%%%%%%%
\section{Unbound Kepler orbits}
\label{app:kepler}
Here, we summarise for our and the readers convenience some well-known relations for non-elliptic Kepler orbits \citep[e.g.][]{Battin1987}. Consider a test particle orbiting a mass $M$ at the origin. Let $\mu\equiv GM$ and $\vec{v}\equiv\dot{\vec{r}}$, then
%\begin{align}
	$\vec{h} = \vec{r}\cross\vec{v}$ and
%	\\
	$\vec{e} = \vec{v}\cross(\vec{r}\cross\vec{v})/\mu - \uvec{r}$
%\end{align}
are its angular-momentum and eccentricity vector, respectively. These are orthogonal and conserved for all Kepler orbits, as is their cross product $\uvec{k}\equiv\uvec{h}\cross\uvec{e}$. Semi-major axis and eccentricity are, respectively,
\begin{align}
	a &= -\frac{\mu}{2E} && = \left[\frac2r - \frac{v^2}\mu\right]^{-1},
	&& && &&
	\\
	e &= |\vec{e}| && = \sqrt{1-h^2/\mu a},
\end{align}
where $E=\tfrac12v^2-\mu/r$ is the (specific) orbital energy.

%%%%%%%%%%%%%%%%%%%%%%%%%%%%%%%%%%%%%%%%%%%%%%%%%%%%%%%
\subsection{Hyperbolic orbits}
\label{app:hyper}
For hyperbolic orbits $E>0$ such that $a<0$ and $e\ge1$ with equality only for $\vec{h}=0$. The radius, position and velocity are
\begin{align}
	\label{rad:hyper:eta}
	r &= -a(e\cosh\eta-1),
	\\
	\label{pos:hyper:eta}
	\vec{r} &= -a (e-\cosh\eta)\uvec{e} - a\sqrt{e^2-1}\sinh\eta\,\uvec{k},
	\\
	\label{vel:hyper:eta}
	\vec{v} &= \frac{\sqrt{-\mu a}}{r}\left(-\sinh\eta\,\uvec{e} + \sqrt{e^2-1}\cosh\eta\,
		\uvec{k}\right),
	\\
	\label{time:hyper:eta}
	\ell &= e\sinh\eta - \eta,
\end{align}
where $\ell=\Omega t$ and $\eta$ are the mean and eccentric anomaly, respectively, while $\Omega^2=-\mu/a^3$. At $t\to\pm\infty$, the orbit asymptotes the straight lines $\vec{r}=\vec{b}_\pm+t\vec{w}_\pm$ with directions
\begin{align}
	\label{eqs:ek:2:wb}
	\uvec{w}_\pm = \left(\sqrt{e^2-1}\,\uvec{k}\mp\uvec{e}\right)/e,
	\qquad
	\uvec{b}_\pm = \left(\sqrt{e^2-1}\,\uvec{e}\pm\uvec{k}\right)/e
\end{align}
and amplitudes
\begin{align}
	\label{eq:w:b}
	w^2 = -\mu/a, \qquad
	b   = -\sqrt{e^2-1}\, a.
\end{align}
From these relations, we may also express semi-major axis and eccentricity through the asymptotes as
\begin{align}
	\label{eq:e,a:b,w}
	a = -\mu/w^2, \qquad
	e = \sqrt{1+b^2w^4/\mu^2}.
\end{align}
The velocity change between the incoming and outgoing asymptotes is
\begin{align}
	\label{eq:delta:v}
	\Delta\vec{v} = -2(w/e)\uvec{e},
\end{align}
which corresponds to a deflection by angle 
\begin{align}
	\label{eq:defl}
	\cos^{-1}(\uvec{w}_\pm\cdot\uvec{w}_\mp) = \cos^{-1}(1-2/e^2).
\end{align}
Hence, $e=\sqrt{2}$, corresponding to $b=-a$, gives a deflection by $90^\circ$. At closest approach $\eta=0$ and
\begin{align}
	\label{eq:closest:r}
	\vec{r}_0 %=\sqrt{\frac{e-1}{e+1}}b\uvec{e}
			= -(e-1)a\uvec{e},
	\qquad
	\vec{v}_0 %=\sqrt{\frac{e+1}{e-1}}w\uvec{k}
			= \sqrt{\frac{e+1}{e-1}} w\uvec{k}.
\end{align}

%%%%%%%%%%%%%%%%%%%%%%%%%%%%%%%%%%%%%%%%%%%%%%%%%%%%%%%
\subsection{Parabolic orbits}
\label{app:para}
For parabolic orbits $E=0$ such that $a=\infty$ and $e=1$, regardless of angular momentum $h$. Radius, position, and velocity are
\begin{align}
	\label{eq:para:r:phi}
	r &= \frac{h^2}{\mu}
		\frac{1}{1+\cos\varphi},
	\\
	\vec{r} &= \frac{h^2}{\mu}
		\frac{\cos\varphi\,\uvec{e}+\sin\varphi\,\uvec{k}}{1+\cos\varphi},
	\\
	\vec{v} &= \frac{\mu}{h}
		\left[-\sin\varphi\,\uvec{e}+(1+\cos\varphi)\uvec{k}\right],
\end{align}
with $\varphi=0$ (true anomaly) at periapse. From $r^2\dot{\varphi}=h$,
\begin{align}
	\label{eq:para:dt}
	\diff t = \frac{h^3}{\mu^2} \frac{\diff\varphi}{(1+\cos\varphi)^2},
\end{align}
which can be integrated to obtain Barker's equation
\begin{align}
	\label{eq:para:t:phi}
	\tan^3\tfrac12\varphi + 3 \tan\tfrac12\varphi = 2B
	\qquad\text{with}\qquad
	B \equiv 3 \frac{\mu^2}{h^3} (t-t_0)
\end{align}
and $t_0$ is the time of periapse. Unlike Kepler's equation~\eqref{time:hyper:eta} for hyperbolic orbits, Barker's equation being cubic has a closed solution:
\begin{align}
	\tan\tfrac12\varphi =
		\left(B+\sqrt{1+B^2}\right)^{1/3}-
		\left(B+\sqrt{1+B^2}\right)^{-1/3}.
\end{align}

%%%%%%%%%%%%%%%%%%%%%%%%%%%%%%%%%%%%%%%%%%%%%%%%%%%%%%%
\label{lastpage}
\end{document}